\documentclass[aoas,preprint]{imsart}
\RequirePackage[OT1]{fontenc}
\RequirePackage{amsthm,amsmath,amsfonts,booktabs}
\RequirePackage[authoryear]{natbib}
\RequirePackage[colorlinks,citecolor=blue,urlcolor=blue]{hyperref}
\usepackage{graphicx}
\usepackage{dcolumn,booktabs,rotating}
\usepackage{multirow}
\usepackage{bm}
\usepackage{enumitem}
\usepackage{comment}
\usepackage{dsfont}
\usepackage{pdflscape}
\usepackage{afterpage}
\newcolumntype{d}[1]{D{.}{.}{#1}}

\pubyear{0000}
\volume{0}
\issue{0}
\firstpage{0}
\lastpage{0}
\arxiv{arXiv:0000.0000}

\startlocaldefs
\numberwithin{equation}{section}
\theoremstyle{plain} \newtheorem{thm}{{\sc Theorem}}
\theoremstyle{plain} \newtheorem{prop}{{\sc Proposition}}
\theoremstyle{plain} 
\theoremstyle{plain} \newtheorem{definition}{{\sc Definition}}
\theoremstyle{plain} \newtheorem{assumption}{{\sc Assumption}}

\theoremstyle{plain}

\theoremstyle{plain} 

\theoremstyle{plain}

\newcommand{\one}{\bm{1}}

\newcommand{\bX}{\bm{X}}
\newcommand{\bZ}{\bm{Z}}

\newcommand{\bS}{\bm{S}}
\newcommand{\bI}{\bm{I}}

\newcommand{\cd}{\stackrel{d}{\rightarrow}}

\newcommand{\ba}{\bm{a}}
\newcommand{\blambda}{\bm{\lambda}}

\newcommand{\bbeta}{\bm{\beta}}
\newcommand{\bgamma}{\bm{\gamma}}
\newcommand{\btheta}{\bm{\theta}}

\DeclareMathOperator\bE{\mathbb E} 
\DeclareMathOperator\bV{\mathbb V} 
\newcommand{\X}{\mathbb{X}}

\newcommand{\C}{\mathbb{C}}

\newcommand{\sumi}{\sum_{i=1}^n}
\newcommand{\sumj}{\sum_{j=1}^J}
\newcommand{\sumk}{\sum_{k=1}^J}

\newcommand{\be}{\begin{eqnarray}}
\newcommand{\ee}{\end{eqnarray}}
\newcommand{\bee}{\begin{eqnarray*}}
\newcommand{\eee}{\end{eqnarray*}}
\newcommand{\bi}{\begin{enumerate}[(i)]}
\newcommand{\ei}{\end{enumerate}}

\endlocaldefs

\begin{document}

\begin{frontmatter}
\title{\small Propensity Score Weighting for Causal Inference with Multiple Treatments}
\runtitle{Propensity Score Weighting with Multiple Treatments}

\begin{aug}
\author{\fnms{Fan} \snm{Li}\ead[label=e1]{fan.f.li@yale.edu}}\and
\author{\fnms{Fan} \snm{Li}\ead[label=e2]{fli@stat.duke.edu}}

\runauthor{F. Li and F. Li}

\affiliation{Yale University and Duke University}

\address{F. Li\\
Department of Biostatistics\\
Yale University\\
135 College St\\
New Haven, Connecticut 06510\\
USA\\
\printead{e1}}

\address{F. Li\\
Department of Statistical Science\\
Duke University\\
122 Old Chemistry Building\\
Durham, North Carolina 27708\\
USA\\
\printead{e2}}
\end{aug}

\begin{abstract}
Causal or unconfounded descriptive comparisons between multiple groups are common in observational studies. Motivated from a racial disparity study in health services research, we propose a unified propensity score weighting framework, the balancing weights, for estimating causal effects with multiple treatments. These weights incorporate the generalized propensity scores to balance the weighted covariate distribution of each treatment group, all weighted toward a common pre-specified target population. The class of balancing weights include several existing approaches such as the inverse probability weights and trimming weights as special cases. Within this framework, we propose a set of target estimands based on linear contrasts. We further develop the generalized overlap weights, constructed as the product of the inverse probability weights and the harmonic mean of the generalized propensity scores. The generalized overlap weighting scheme corresponds to the target population with the most overlap in covariates across the multiple treatments. These weights are bounded and thus bypass the problem of extreme propensities. We show that the generalized overlap weights minimize the total asymptotic variance of the moment weighting estimators for the pairwise contrasts within the class of balancing weights. We consider two balance check criteria and propose a new sandwich variance estimator for estimating the causal effects with generalized overlap weights. We apply these methods to study the racial disparities in medical expenditure between several racial groups using the 2009 Medical Expenditure Panel Survey (MEPS) data. Simulations were carried out to compare with existing methods.
\end{abstract}


\end{frontmatter}

\section{Introduction}\label{sec:intro}
Propensity score weighting is a common method for balancing covariates and estimating treatment effects in causal inference \citep{Rosenbaum1983}. It is also applicable to unconfounded non-causal comparisons such as racial disparities studies \cite[e.g.][]{McGuire2006,Cook2009}. There is a vast literature on propensity score weighting with binary treatments; see, for example, a recent review by \cite{ding2018causal}. This paper focuses on propensity score weighting strategies for multiple group comparisons, which have become increasingly common in practice. For example, in comparative effectiveness research, the interest often lies in comparing the effectiveness of several medical treatments; in health service research, the interest often lies in examining the disparities in health care utilization between more than two races or ethnicities \citep{Zaslavsky2005}.

For multiple group comparisons, \citet{Imbens2000} extended the classic results of \citet{Rosenbaum1983} and developed the generalized propensity score method; the key insight is that the scalar generalized propensity score of each treatment level can be exploited to separately estimate the average potential outcomes in that group. With the generalized propensity score device, matching and subclassification strategies have been discussed extensively; see, for instance, \citet{Lechner2002,Zanutto2005,Rassen2013,Yang2016, Lopez2017b}. With the weighting strategy, the existing methods for multiple-group comparisons have largely focused on the pairwise average treatment effect (ATE), based on the inverse probability weighting (IPW) \citep{Feng2012,McCaffrey2013}. However, observational studies often rely on convenience samples, which does not necessarily represent a population of scientific meaning. In such cases, the automatic focus on ATE may be questionable because it is not clear what target population the causal conclusion is applicable to. Meanwhile, multiple treatments exacerbate the overlap issues as different treatments may be applicable only to certain subpopulations, and the ATE may correspond to an infeasible intervention. Regardless of the number of treatment levels, extreme propensity scores close to zero or one will likely result in bias and excessive variance of the IPW estimators \citep{LiThomasLi2018}. \citet{Crump2009} proposed an optimal trimming procedure that focuses on regions with good overlap and thus improves the efficiency of the IPW estimator for binary treatments; \citet{Yang2016} extended the trimming rule to more than two treatments. Though easy to implement, propensity trimming often leads to an ambiguous target population and may discard a large number of units.

In this article, we propose a unified propensity score weighting framework for causal inference with multiple treatments. Specifically, we generalize the balancing weights framework for binary treatments \citep{LiMorganZaslavsky2018} to balance the distribution of covariates from multiple treatment groups according to a pre-specified target population. Within this framework, we propose a set of target estimands based on linear contrasts. We further develop the generalized overlap weights, constructed as the product of the inverse probability weights and the harmonic mean of the generalized propensity scores. The generalized overlap weights focus on the subpopulation with substantial probabilities to be assigned to all treatments. This target population aligns with the spirit of randomized clinical trials by emphasizing patients at clinical equipoise, and is thus of natural relevance to medical and policy studies. Under mild conditions, we show that the generalized overlap weights minimize the total asymptotic variance of the moment estimators for the pairwise contrasts within the class of balancing weights. These new weights are strictly bounded between zero and one, and thus automatically bypass the issue of extreme propensity scores.

Our methodological innovation is motivated by an application to racial disparities in medical expenditure. Identifying and tracking racial disparities in health care utilization represents a crucial step in developing health care policy and allocating health services resources. The \emph{Unequal Treatment} report from the Institute of Medicine (IOM) defined health care disparity as the difference in treatment provided to social groups that is not justified by health status or treatment preference of the patient \citep{IOM2003}. Therefore, adjusting for the health status variables across different racial groups is necessary for producing interpretable disparity estimates concordant with the IOM definition. In this sense, these descriptive comparisons share the same nature with causal comparisons with respect to confounding control, and indeed propensity score methods have been widely used in health care disparity studies \citep{Cook2012}. One particular challenge is that the IOM definition of disparity includes racial differences in utilization mediated through factors other than health status and preference, such as many social factors \citep{McGuire2006}. Accordingly, a number of methods have been developed to account for the socioeconomic status variables in the propensity score analysis of racial disparities in health services \citep[e.g.][]{McGuire2006, Cook2009}. In this paper, we combine one such method---the rank-and-replace adjustment---with the proposed generalized overlap weights to track racial disparities in medical expenditure between Whites, Blacks, Hispanics and Asians. This is in contrast to most existing racial disparity studies, which conducted separate comparisons of each White-minority pair \citep{Cook2010}.

The remainder of this article is organized as follows. Section \ref{sec:balancweights} introduces the general framework of balancing weights. In Section \ref{sec:ow}, we propose the generalized overlap weights for pairwise comparisons with multiple treatments, discuss balance check criteria and variance estimation. In Section \ref{sec:app}, we reanalyze the Medical Expenditure Panel Survey data and study the racial disparities in medical expenditure between several racial groups. Section \ref{sec:sim} carries out simulations to examine the operating characteristics of the proposed method and compare with existing methods. Section \ref{sec:dis} concludes.

\section{Balancing Weights for Multiple Treatments} \label{sec:balancweights}
\subsection{Basic Setup}\label{sec:setup}
We consider a sample of $n$ units, each belonging to one of $J\geq 3$ groups for which covariate-balanced comparisons are of interest.  Let $Z_i\in \mathbb{Z}=\{1,\ldots,J\}$ denote the treatment group membership, and $D_{ij}=\mathds{1}\{Z_i=j\}$ the indicator of receiving treatment level $j$. For each unit, we observe an outcome $Y_i$ and a set of $p$ pre-treatment covariates $\bX_i=(X_{i1},...,X_{ip})'$. For $J\geq 3$ treatments, \citet{Imbens2000} defined the generalized propensity score, as follows.
\begin{definition}(Generalized Propensity Scores)
The generalized propensity score is the conditional probability of being assigned to each group given the covariates:
\bee
e_j(\bX)=\Pr(Z=j|\bX), \quad j\in \mathbb{Z}.
\eee
\end{definition}

By definition, the sum-to-unity restriction $\sum_{j=1}^J e_j(\bX)=1$ holds for all $\bX$ in support $\X$, and hence each unit's propensity can be uniquely characterized by $J-1$ scalar scores. Under the Stable Unit Treatment Value Assumption (SUTVA), each unit has a potential outcome $Y_{i}(j)$ mapped to each treatment level $j\in\mathbb{Z}$, among which, only the one corresponding to the received treatment, $Y_i=Y_i(Z_i)$, is observed. To proceed, we make the following two standard assumptions.

\begin{assumption}\label{asp:unconfound}(Weak Unconfoundedness) The assignment is weakly unconfounded if
\bee
Y(j)\perp \mathds{1}\{Z=j\} |\bX,~~~\forall~j\in \mathbb{Z}.
\eee
\end{assumption}
\begin{assumption}\label{asp:overlap}(Overlap) For all $\bX\in\X$ and all group $j$, the probability of being assignment to any treatment group is bounded away from zero:
\bee
e_j(\bX)>0,~~~\forall~\bX\in\mathbb{X},~j\in\mathbb{Z}.
\eee
\end{assumption}

Assumption \ref{asp:unconfound} imposes unconfoundedness separately for each level of the treatment, and is sufficient for identification of the population-level estimand \citep{Imbens2000}. This assumption implies that the potential outcome $Y(j)$ is independent of the assignment indicator $\mathds{1}\{Z=j\}$, conditional on the scalar generalized propensity score $e_j(\bX)$. In other words, adjusting for the scalar score is sufficient to remove the bias in estimating the average value of $Y(j)$ over the target population. Assumption \ref{asp:overlap} restricts the study population to the covariate space where each unit has non-zero probability to receive any treatment. 

To elaborate, we define the conditional expected potential outcomes in group $j$ as $m_j(\bX)=\bE[Y(j)|\bX]$. Under Assumption \ref{asp:unconfound}, we have $m_j(\bX)=\bE[Y|Z=j,\bX]$, which is estimable from the observed data. As previously mentioned, the propensity score methods are also applicable to unconfounded descriptive (non-causal) comparisons where the group membership is a non-manipulable state, such as different races and different years. In these cases, a common objective is to compare the expected observed outcomes, $m_j(\bX)=\bE[Y|Z=j,\bX]$; for example, when $J=2$, \citet{LiZaslavskyLandrum2013} defined the contrast between $m_1(\bX)$ and $m_2(\bX)$ averaged over a population as the \emph{average controlled difference} (ACD). For simplicity, henceforth we use the nomenclature of causal inference to generically refer to both causal and unconfounded descriptive settings, but emphasize that the methods developed here are applicable to both.

\subsection{Balancing Weights}\label{sec:bw}
Assume the marginal density of the covariates, $f(\bX)$, exists, with respect to a base measure $\mu$. In causal studies, the interest is on the average effects of units in a target population, whose density (up to a normalizing constant) we represent by $g(\bX)=f(\bX)h(\bX)$, with $h(\bX)$ being a pre-specified function of covariates, which we refer to as a tilting function. We first define the expectation of the potential outcomes over the target population $g(\bX)$:
\be\label{eq:meanpo}
m_j^h \equiv \frac{\int_{\X} m_{j}(\bX)f(\bX)h(\bX)\mu(d\bX)}{\int_{\X} f(\bX)h(\bX)\mu(d\bX)}.
\ee
Then we characterize a class of additive estimands as a linear combination of the above expectations, with coefficients $\ba=(a_1,\cdots, a_J)'$:
\be\label{eq:estimand}
\tau^{h}(\ba)\equiv \sumj  a_j m_j^h.
\ee
The causal estimand $\tau^{h}(\ba)$ generalizes the definition of weighted average treatment effect (WATE) in binary treatments \citep{Hirano2003} where $J=2$ and $\ba=(1,-1)$. As will be seen in due course,  $\tau^{h}(\ba)$ includes several existing causal estimands as special cases.

We next define the class of balancing weights. Let $f_j(\bX)=f(\bX|Z=j)$ be the density of $\bX$ in the $j$th group over its support $\X_j$, we have $f_j(\bX)\propto f(\bX)e_j(\bX)$. Given any pre-specified function $h$, we can weight the group-specific density $f_j(\bX)$ to the target population using the following weights, proportional up to a normalizing constant:
\be\label{eq:bweight}
w_j(\bX) \propto  \frac{f(\bX)h(\bX)}{f(\bX)e_j(\bX)} =\frac{h(\bX)}{e_j(\bX)}, \quad \forall ~ j\in\mathbb{Z}.
\ee
It is straightforward to show that the class of weights defined in (\ref{eq:bweight}) balance the weighted distributions of the covariates across $J$ comparison groups:
\begin{equation}\label{eq:balance}
f_j(\bX)w_j(\bX)= f(\bX)h(\bX),  \quad \forall ~ j\in\mathbb{Z}.
\end{equation}

To apply the above framework, a key is to specify the coefficients $\ba$ and the tilting function $h$, with the former defining the causal contrast and the latter representing the target population. We focus on the case of multiple nominal treatments, where the scientific interest usually lies in pairwise comparisons. More specifically, the choice of $\ba$ is contained in the finite set $\mathbb{S}=\{\blambda_{j,j'}=\blambda_j-\blambda_{j'}:j<j'\}$, where $\blambda_j$ is the $J\times 1$ unit vector with one at the $j$th position and zero everywhere else. In principle, the tilting function $h$ can take any form, each leading to a unique type of balancing weights; statistical, scientific and policy considerations all play into the specification of $h$. We illustrate specifications of $\ba$ and $h$ (up to a normalizing constant) by connecting the general definition (\ref{eq:estimand}) with existing estimands in the causal inference literature.

When $h(\bX)=1$, the target population $f(\bX)$ is the combined population from all groups and the weights become the standard inverse probability weights, $\{1/e_j(\bX),~j\in\mathbb{Z}\}$; the target estimand is the pairwise ATE as in \citet{Feng2012}. When $h(\bX)=e_{j'}(\bX)$, the target population is the subpopulation receiving treatment $Z=j'$, and the weights, $\{e_{j'}(\bX)/e_j(\bX),~j\in\mathbb{Z}\}$, are designed to estimate the average treatment effect for the treated (ATT). Define
$$\underline{e}_j=\max_{1\leq l\leq J} \{\min_{\bX\in \X_l}\{e_j(\bX)\} \},~~~
\bar{e}_j=\min_{1\leq l\leq J} \{\max_{\bX\in \X_l}\{e_j(\bX)\} \},$$
and an eligibility function $E_j(\bX)=\mathds{1}\{\underline{e}_j\leq e_j(\bX)\leq \bar{e}_j\}$ for all $j\in\mathbb{Z}$. When $h(\bX)=e_{j'}(\bX)\prod_{j=1}^J E_j(\bX)$, the target population is the subpopulation receiving treatment $Z=j'$ but remaining eligible for all other treatments \citep{Lopez2017b}. Similar eligibility functions were used earlier by \citet{VanderLaan2007} and \citet{Moore2012} to develop improved causal models with time-varying treatments. Further, define a threshold $\alpha$ as the largest value such that
\be\label{eq:trim}
\alpha\leq \frac{2\bE\left[\sumj 1/e_j(\bX) | \sumj 1/e_j(\bX)\leq \alpha \right]}
{\text{Pr}\left(\sumj 1/e_j(\bX)\leq \alpha\right)}.
\ee
When $h(\bX)=\mathds{1}\{\bX\in \C\}$ with $\C=\{\bX\in\X|\sumj 1/e_j(\bX)\leq \alpha\}$, the target population is characterized by the subpopulation $\C$, and the inverse probability weights are formulated after applying the optimal trimming rule \citep{Yang2016}. Finally, when $h(\bX)=\min_{1\leq k\leq J}\{e_k(\bX)\}$, one arrives at the generalized matching weights \citep{Yoshida2017}---an extension of the matching weights of \citet{LiGreene13} to multiple treatments. Such an approach represents a weighting analogue to exact matching and the causal comparisons are made for the matched population. When $h(\bX)=\bV(\mathds{1}\{Z=j'\}|\bX)=e_{j'}(\bX)\{1-e_{j'}(\bX)\}$, the target estimand becomes the $j'$th variance-weighted average treatment effect studied by \citet{Robins2008}, who also proposed efficient and flexible estimators based on higher-order influence functions. Finally, one could choose indicator functions for $h$ that directly involves covariates of a subpopulation of interest, such as a specific gender or a range of age. Table \ref{tb:example} summarizes the above special cases.

\begin{table}[htbp]
\centering
\caption{Examples of balancing weights and target populations for making pairwise comparisons with different tilting functions.}\label{tb:example}\vspace{0.1in}
\begin{tabular}{lll}
\toprule
 Target Population & Tilting Function $h(\bX)$ & Weights $\{w_j(\bX),~j\in\mathbb{Z}\}$\\\midrule
Combined & 1 & $\{1/e_j(\bX),~j\in\mathbb{Z}\}$\\
$j'$th Treated & $e_{j'}(\bX)$ & $\{e_{j'}(\bX)/e_j(\bX),~j\in\mathbb{Z}\}$ \\
$j'$th Treated (restricted) & $e_{j'}(\bX)\prod_{j=1}^J E_j(\bX)$ & $\{e_{j'}(\bX)\prod_{j=1}^J E_j(\bX)/e_j(\bX),~j\in\mathbb{Z}\}$\\
Trimming & $\mathds{1}\{\bX\in \C\}$ & $\{\mathds{1}\{\bX\in \C\}/e_j(\bX),~j\in\mathbb{Z}\}$\\
Generalized Matching & $\min_{1\leq k\leq J}\{e_k(\bX)\}$ & $\{\min_{1\leq k\leq J}\{e_k(\bX)\}/e_j(\bX),~j\in\mathbb{Z}\}$\\
$j$'th Variance-Weighted & $e_{j'}(\bX)\{1-e_{j'}(\bX)\}$ & $\{e_{j'}(\bX)\{1-e_{j'}(\bX)\}/e_j(\bX),~j\in\mathbb{Z}\}$\\
Generalized Overlap & $(\sumk  1/e_k(\bX))^{-1}$ & $\{(\sumk  1/e_k(\bX))^{-1}/e_j(\bX),~j\in\mathbb{Z}\}$\\
\bottomrule
\end{tabular}
\end{table}

When the treatment levels are ordered categories, target estimands may differ from the pairwise comparisons and require different choice of $\ba$. For instance, one may be interested in the quadratic contrasts between unit increases in the treatment level, namely $(m_{j+1}^h-m_j^h)-(m_j^h-m_{j-1}^h)$. In other cases, one may estimate the weighted average of unit increase in the treatment level, $\sum_{j=1}^{J-1}\pi_j (m^h_{j+1}-m^h_j)$, or the accumulative effect of the maximum treatment, $m^h_J-m^h_1$. For the disparity study in Section \ref{sec:app}, the multiple racial groups are unordered categories. For this reason, we mainly focus on multiple nominal groups, but note that the general framework of balancing weights remains applicable to multiple ordinal groups.

\subsection{Large-sample Properties of Moment Estimators} \label{sec:asymp}
For any pre-specified vector $\ba$ and tilting function $h$, we could first use the plug-in sample moment estimator to obtain the expectation of the potential outcomes among the target population
\be\label{eq:samplemeanpo}
\hat{m}^h_j=\frac{\sum_{i=1}^n D_{ij}Y_i w_j(\bX_i) }{\sum_{i=1}^n D_{ij}w_j(\bX_i)},
\ee
and then estimate $\tau^h(\ba)$ by a linear combination, $\hat{\tau}^{h}(\ba)=\sum_{j=1}^J  a_j \hat{m}_j^h$, where the sum is over a sample drawn from density $f(\bX)$. Below we establish three large-sample results of $\hat{\tau}^h(\ba)$; the proofs are given in Section B of the Supplementary Material \citep{LiLiSupp2019}.
\begin{prop}\label{prop1}
Given any $h$ and $\ba$, $\hat{\tau}^h(\ba)$ is a consistent estimator of $\tau^h(\ba)$.
\end{prop}

Denote the collection of treatment assignment $\underline{\bZ}=\{Z_1,\ldots,Z_n\}$ and covariate design points $\underline{\bX}=\{\bX_1,\ldots,\bX_n\}$. The next two results concern the variance of the sample estimator, which is decomposed as
\bee
\bV[\hat{\tau}^h(\ba)]=\bE_{\underline{\bZ},\underline{\bX}}\bV[\hat{\tau}^h(\ba)|\underline{\bZ},\underline{\bX}]+\bV_{\underline{\bZ},\underline{\bX}}\bE[\hat{\tau}^h(\ba)|\underline{\bZ},\underline{\bX}].
\eee
The first term is the variation due to residual variance in $\hat{\tau}^h(\ba)$ conditional on the design points. The second term arises from the dependence of the expectation of the plug-in estimator on the sample, and estimating it involves the outcome model (associations between $Y(j)$ and $\bX$). As individual variation is typically much larger than conditional mean variation, the benefit of further optimizing the weights by a preliminary look at the outcomes, which mixes the design and analysis, would usually not justify the risk of biasing model specification to attain desired results \citep{Imbens2004}. Hence, we focus on the first term.

\begin{prop}\label{prop2}
Given $\ba$, suppose the family of residual variances \\ $\{\bV[\hat{\tau}^h(\ba)|\underline{\bZ},\underline{\bX}],n\geq 1\}$ is uniformly integrable. Then the expectation of the conditional variance converges
\bee
n \cdot\bE_{\underline{\bZ},\underline{\bX}}\bV [\hat{\tau}^h(\ba)|\underline{\bZ},\underline{\bX}]\rightarrow Q(\ba,h)
\equiv \int_{\X} \Big(\sum_{j=1}^J a_j^2v_j(\bX)/e_j(\bX)\Big) h^2(\bX)f(\bX)\mu(d\bX)/C_h^2,
\eee
where $v_j(\bX)=\bV[Y(j)|\bX]$ and $C_h\equiv \int_{\X} h(\bX)f(\bX)\mu(d\bX)$ is a constant.
\end{prop}

When the residual variance of the potential outcome is homoscedastic across all groups such that $v_j(\bX)=v$, then the limit $Q(\ba,h)$ can further simplify and the following result holds.

\begin{prop}\label{prop:hopt}
Under homoscedasticity, the function
\bee
\tilde{h}(\bX)\propto \frac{1}{\sumj a_j^2/e_j(\bX)}
\eee
gives the smallest asymptotic variance for the moment estimator $\hat{\tau}^h(\ba)$ among all $h$'s, and $\min_h Q(\ba,h)= v/C_{\tilde{h}}$.
\end{prop}

A more general result of Proposition \ref{prop:hopt} can be obtained under heteroscedasticity. In that case, the optimal tilting function,
\bee
\tilde{h}(\bX)\propto \frac{1}{\sumj a_j^2v_j(\bX)/e_j(\bX)},
\eee
explicitly depends on the residual variances of the potential outcomes. Although estimates of $v_j(\bX)$ can be obtained by outcome regression modeling in the analysis stage, it is rarely the case that accurate prior information is available in the design stage. Therefore, such a tilting function is difficult to specify for design purposes and may find limited use without peeking at the outcomes. For such considerations, we motivate the generalized overlap weights in Section \ref{sec:ow} under homoscedasticity. These asymptotic results generalize those for binary treatments in \citet{LiMorganZaslavsky2018}; they also extend the asymptotic results on propensity score trimming in \citet{Crump2009} and \citet{Yang2016}, who have similarly assumed homoscedasticity but restricted the class of tilting functions to indicator functions.

\section{Generalized Overlap Weighting for Pairwise Comparisons}\label{sec:ow}
\subsection{The Generalized Overlap Weights}\label{sec:gow}
For nominal treatments, scientific interest often lies in comparing outcomes between each pair of treatment groups in a common target population. In this case, as $\ba\in \mathbb{S}$, we propose to choose the tilting function $h$ that minimizes the total asymptotic variance of the sample estimators for all pairwise comparisons; in other words, the objective function is
\bee
\sum_{j< j'}Q(\blambda_{j,j'},h)\propto Q(\one_{J},h),
\eee
where $\one_{J}$ is the $J\times 1$ vector of ones. According to Proposition \ref{prop:hopt}, the function $h(\bX)=(\sumj 1/e_j(\bX))^{-1}$---the harmonic mean of the generalized propensity scores---minimizes $Q(\one_{J},h)$ among all choices of $h$. Based on this optimal tilting function $h$, we define the generalized overlap weights for $j=1,...,J$:
\bee
w_j(\bX) \propto \frac{1/e_j(\bX)}{\sumk  1/e_k(\bX) }.
\eee
For binary treatments ($J=2$), the generalized overlap weights reduce to the overlap weights in \citet{LiMorganZaslavsky2018}, namely the propensity of assignment to the other group: $w_1(\bX)\propto 1-e_1(\bX)=e_2(\bX)$, $w_2(\bX)\propto 1-e_2(\bX)=e_1(\bX)$.

\begin{figure}[!h]
\centering
\includegraphics[scale=0.6]{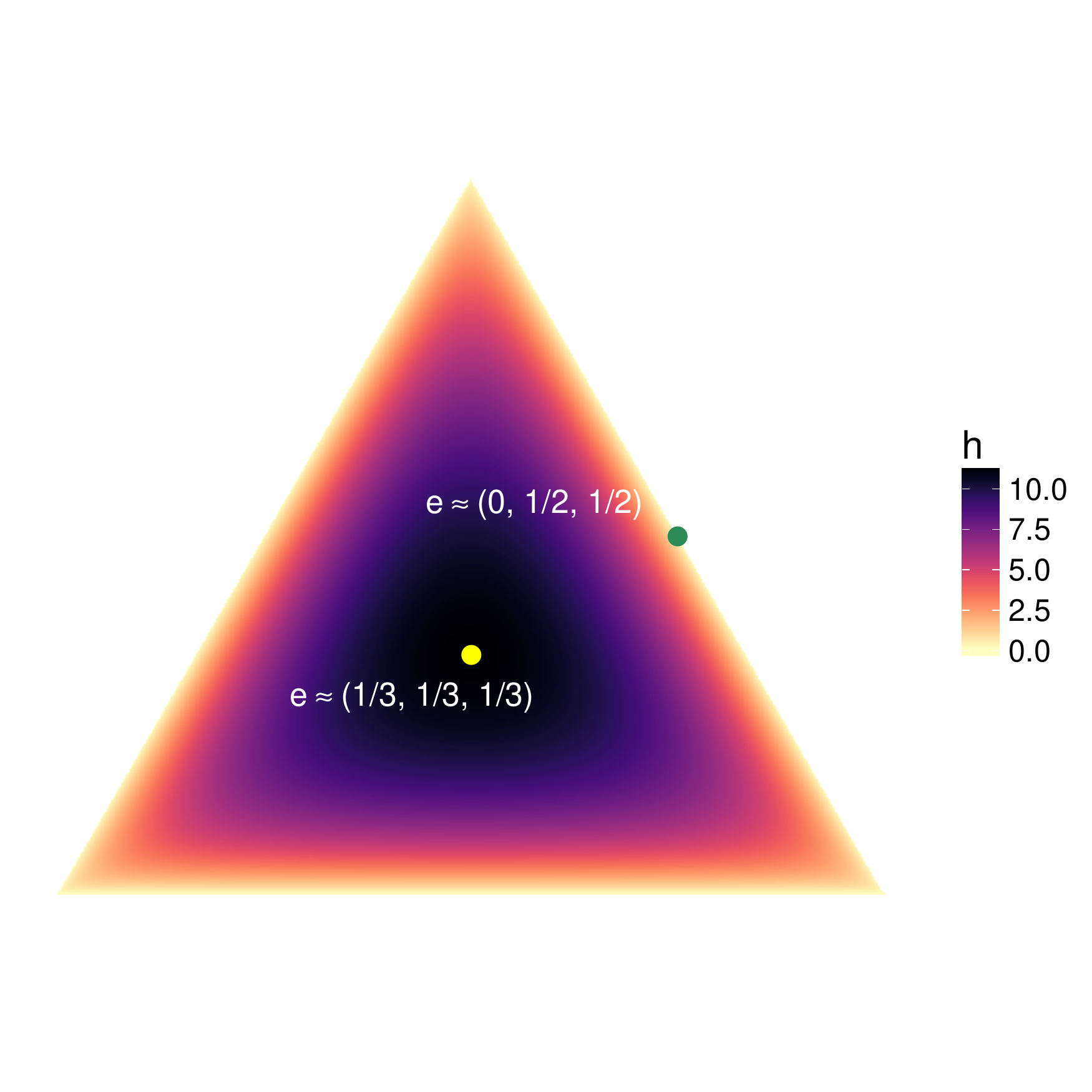}
\caption{Ternary plot of optimal $h$ (up to a proportionality constant) as a function of the generalized propensity score vector with $J=3$ treatments. Each point in the triangular plane represents a unit with certain values of the generalized propensity scores. The value of each generalized propensity score is proportional to the orthogonal distance from that point to each edge. It is evident that the new weighting scheme emphasizes the centroid region with good overlap, e.g., units with $e(\bX)\approx (1/3, 1/3, 1/3)$, and smoothly down-weights the edges, e.g., units with $e(\bX)\approx (0, 1/2, 1/2)$.}
\label{fig:visualize}
\end{figure}

The maximum of the harmonic mean function $h$ is attained when $e_j(\bX)=1/J$ for all $j$, that is, when the units have the same propensity to each of the treatments. Heuristically, the tilting function $h$ gives the most relative weight to the covariate regions in which none of the propensities are close to zero. While it is generally difficult to visualize the optimal $h$ in higher dimensions, we could do so with $J=3$ treatments. Figure \ref{fig:visualize} provides a ternary plot of $h$ when $J=3$. It is clear that the optimal tilting function gives the most relative weight to the covariate regions in which none of the propensities are close to zero, and down-weights the region where there is lack of overlap in at least one dimension. Therefore, we can interpret the corresponding target population to be the subpopulation with the most overlap in covariates among all groups, and term the target estimand as the pairwise average treatment effect among the overlap population (ATO). As the overlap population tilts $f(\bX)$ most heavily toward equipoise, it is naturally of policy and clinical relevance. Especially for clinical practice, this target population aligns with the spirit of randomized studies and emphasizes patients with clinical equipoise, whose treatment decisions remain unclear and thus for whom comparative information is most needed. Analogously, in descriptive studies for racial disparities, the overlap population represents individuals with most similarity in observed health-related characteristics, based on whom subsequent policy interventions on health care utilization become most meaningful.

Besides asymptotic efficiency, the generalized overlap weights have several attractive features. First, the harmonic mean function $h$ is strictly bounded
\bee
0< \min_{1\leq k\leq J}\{e_k(\bX)\}/J \leq h(\bX) \leq \min_{1\leq k\leq J}\{e_k(\bX)\} < 1,
\eee
and thus the weighting scheme is robust to extreme weights, in contrast to IPW. Second, the target population defined by the generalized overlap weights is adaptive to the covariate distributions among the $J$ comparison groups. For example, when the propensity of assignment to treatment $j$ is small compared to others so that $e_j(\bX)\approx 0$, the tilting function
\bee
h(\bX)\propto \prod_{l=1}^J e_l(\bX)/\sum_{k=1}^J \prod_{l\neq k} e_l(\bX)\approx \prod_{l=1}^J e_l(\bX)/\prod_{l\neq j} e_l(\bX)=e_j(\bX),
\eee
suggesting that the target population is similar to the $j$th treatment group and the associated estimand approximates the ATT. On the other hand, if the treatment groups are almost balanced in size and covariate distribution so that $e_j(\bX)\approx 1/J$ for all $j$, we have $h(\bX)\propto 1$ and the target estimand approximates the pairwise ATE. Arguably this adaptiveness enables the generalized overlap weighting scheme to define a scientific question that may be best answered nonparametrically by the available data at hand. Finally, the generalized matching weights \citep{Yoshida2017}---defined by $h(\bX)=\min_{1\leq j\leq J}\{e_j(\bX)\}$---share some of the above advantages, but these weights are not asymptotically efficient and are non-smooth, which renders the variance calculation more complex.

\subsection{Estimate Generalized Propensity Scores and Balance Check}\label{sec:BalanceCheck}
In practice, usually the propensity scores are not known and must be estimated from the data. For multiple nominal treatments, the generalized propensity scores are frequently modeled by a multinomial logistic regression,
\be
e_1(\bX_i)&=&\frac{1}{1+\sum_{k=2}^{J}\exp(\alpha_k+\bX_i^T\bbeta_k)}, \nonumber\\
e_j(\bX_i)&=&\frac{\exp(\alpha_j+\bX_i^T\bbeta_j)}{1+\sum_{k=2}^{J}\exp(\alpha_k+\bX_i^T\bbeta_k)},~~~~j=2,\ldots,J,
\label{eq:multinomial}
\ee
where the covariate vector $\bX$ are allowed to contain higher-order moments, splines and interactions. Model parameters $\btheta=(\alpha_2,\ldots,\alpha_J,\bbeta^T_2,\ldots,\bbeta^T_J)^T$ can be estimated by standard maximum likelihood, from which we obtain the estimated propensity scores. To assess the fit of the propensity score model, we check the weighted covariate balance in the target population. We consider two ways for balance check motivated by the population balancing constraint (\ref{eq:balance}). First, constraint (\ref{eq:balance}) implies the weighted covariate balance between each group and the target population. Therefore, we inspect, for each treatment level, the weighted covariate mean deviation from that of the target population. Specifically, we define $\bar{X}_j={\sum_{i=1}^n D_{ij}X_i w_j(\bX_i)}\Big/{\sum_{i=1}^n D_{ij} w_j(\bX_i)}$ as the weighted mean of covariate $X$ from the $j$th group and $S^2_{X,j}$ as the unweighted variance. Further, we define $\bar{X}_{p}={\sum_{i=1}^n X_i h(\bX_i)}\Big/{\sum_{i=1}^n h(\bX_i)}$ as the average value of covariate $X$ in the target population and $S^2_{X}=J^{-1}\sumj S^2_{X,j}$ as the averaged unweighted variance. The population standardized difference (PSD) is then defined for each covariate and each treatment level as $\text{PSD}_j=|\bar{X}_j-\bar{X}_{p}|/S_X$. Similar to \citet{McCaffrey2013}, we then use $\max_j|\text{PSD}_j|$ as the balance metric for each covariate $X$ and inspect the adequacy of the propensity score model. If a covariate is not well balanced in one group, interaction terms of that variable with other variables can be added to the model, and the new model is re-fit and re-evaluated until balance is deemed satisfactory. On the other hand, the population balance constraint (\ref{eq:balance}) also implies pairwise balance $f_j(\bX)w_j(\bX)=f_{j'}(\bX)w_{j'}(\bX)$ for all $j\neq j'$, and so we could alternatively assess balance by checking the pairwise absolute standardized differences (ASD), $\text{ASD}_{j,j'}=|\bar{X}_j-\bar{X}_{j'}|/S_X$. The balance metric for each covariate can then be similarly specified as $\max_{j<j'}|\text{ASD}_{j,j'}|$.

Finally, a special property of the overlap weights with binary treatments is exact balance, that is, when the propensity scores are estimated from a logistic model, the standardized difference of all the covariates entering the propensity model is zero, i.e, $\text{ASD}_{1,2}=0$ for $J=2$ \citep[][Theorem 3]{LiMorganZaslavsky2018}. However, this exact balance property is due to the happenstance that the logistic score equations exploit the covariate-balancing moment conditions, and does not directly extend to the generalized overlap weights with $J\geq 3$ when the propensity score is estimated by a multinomial logistic model. Therefore, we still recommend the conventional iterative fitting-checking procedure to improve the propensity model.

\subsection{Variance Estimation} \label{sec:variance}
The asymptotic variance results in Section \ref{sec:asymp} are not directly useful for calculating the sample variance of $\hat{\tau}^h(\blambda_{j,j'})$ in practice because the $v_j(\bX)$'s are not known. Moreover, one has to account for the additional uncertainty in estimating the propensities in the variance estimation. Here we derive an empirical sandwich variance estimator \citep{Stefanski2002} that accounts for the uncertainty in estimating the generalized overlap weights from the multinomial logistic model \eqref{eq:multinomial}. We provide the following theorem to motivate the closed-variance calculation for the pairwise ATO estimates. The proof is given Section C of the Supplementary Material \citep{LiLiSupp2019}.
\begin{thm}\label{thm:var}
Under standard regularity conditions, when the generalized propensity scores are estimated by multinomial logistic regression (\ref{eq:multinomial}), the resulting ATO estimator between groups $j$ and $j'$ is asymptotically normal
\bee
\sqrt{n}\{\hat{\tau}^h(\blambda_{j,j'})-\tau^h(\blambda_{j,j'})\}\cd \mathcal{N}\left(0,{\bE\left\{\psi_{ij}-\psi_{ij'}\right\}^2}/{[\bE\{h(\bX)\}]^2}\right),
\eee
where
\bee
\psi_{ij}=D_{ij}(Y_i-m^h_j)w_j(\bX_i)+\bE\left\{D_{ij}(Y_i-m^h_j)\frac{\partial}{\partial\btheta^T}w_j(\bX_i)\right\}\bI_{\btheta\btheta}^{-1} \bS_{\btheta,i},
\eee
and $\bS_{\btheta,i}$, $\bI_{\btheta\btheta}$ are the individual score and information matrix of $\btheta$, respectively.
\end{thm}

Theorem \ref{thm:var} suggests the following consistent variance estimator. Denote $\hat{\btheta}$, $\hat{\bS}_{\btheta,i}$, $\hat{\bI}_{\btheta\btheta}$ as the maximum likelihood estimator of $\btheta$, the plug-in consistent estimators for the individual score and information matrix, the variance estimator for the estimated ATO is expressed by
\be\label{eq:sand}
\hat{\bV}[\hat{\tau}^h(\blambda_{j,j'})]=\frac{\sumi \left(\hat{\psi}_{ij}-\hat{\psi}_{ij'}\right)^2}
{\left[\sumi \{\sumk 1/\hat{e}_k(\bX_i)\}^{-1}\right]^2},
\ee
where
\bee
\hat{\psi}_{ij}=
D_{ij}(Y_i-\hat{m}^{h}_j)w_j(\bX_i;\hat{\btheta})+
\left\{\frac{1}{n}\sumi D_{ij}(Y_i-\hat{m}^{h}_j)
\frac{\partial}{\partial\btheta^T}w_j(\bX_i;\hat{\btheta})\right\}\hat{\bI}_{\btheta\btheta}^{-1} \hat{\bS}_{\btheta,i}.
\eee

The true generalized propensity score is generally unknown in applications and will be substituted by its sample analogue. \citet{Hirano2003} suggested that a consistent estimator of the propensity score leads to more efficient estimation of the WATE with binary treatments than the true propensity score. Our derivation of the variance estimator re-interprets their findings in the context of multiple treatments. Specifically, with a consistent estimator for the generalized propensity score, the influence function for estimating $m_j^h$, $\psi_{ij}/\bE\{h(\bX)\}$, can be viewed as the residual of $D_{ij}(Y_i-m^h_j)w_j(\bX_i)/\bE\{h(\bX)\}$---the influence function for estimating $m_j^h$ using the true propensity score---after projecting it onto the nuisance tangent space of $\btheta$. Therefore, the efficiency implications from \citet{Hirano2003} carry over to our pairwise comparisons emphasizing the overlap population.

\section{Application to Racial Disparities in Medical Expenditure}\label{sec:app}
\subsection{The Data}\label{sec:data}
Our application is based on the 2009 Medical Expenditure Panel Survey (MEPS) data. The sample contains health information, socioeconomic status (SES) and total health care expenditure for four racial groups with adult aged at least 18 years: 9830 non-Hispanic Whites, 1446 Asians, 4020 Blacks, 5150 Hispanics. We are interested in estimating the health care disparity in the yearly total health care expenditure, after controlling for the differences due to patient health status, i.e., variables reflecting clinical appropriateness and need. Using the MEPS data, \citet{Cook2010} estimated the racial disparities between each White-minority pair. One potential limitation of such separate binary comparisons is the non-transitivity among the pairwise estimates, as each comparison may be made for a different target population (see Section A of the Supplementary Material \citep{LiLiSupp2019} for a detailed discussion on transitivity). Here we focus on the simultaneous multiple-group comparisons by defining a common target population.

\begin{figure}[!h]
\centering
\includegraphics[scale=0.4]{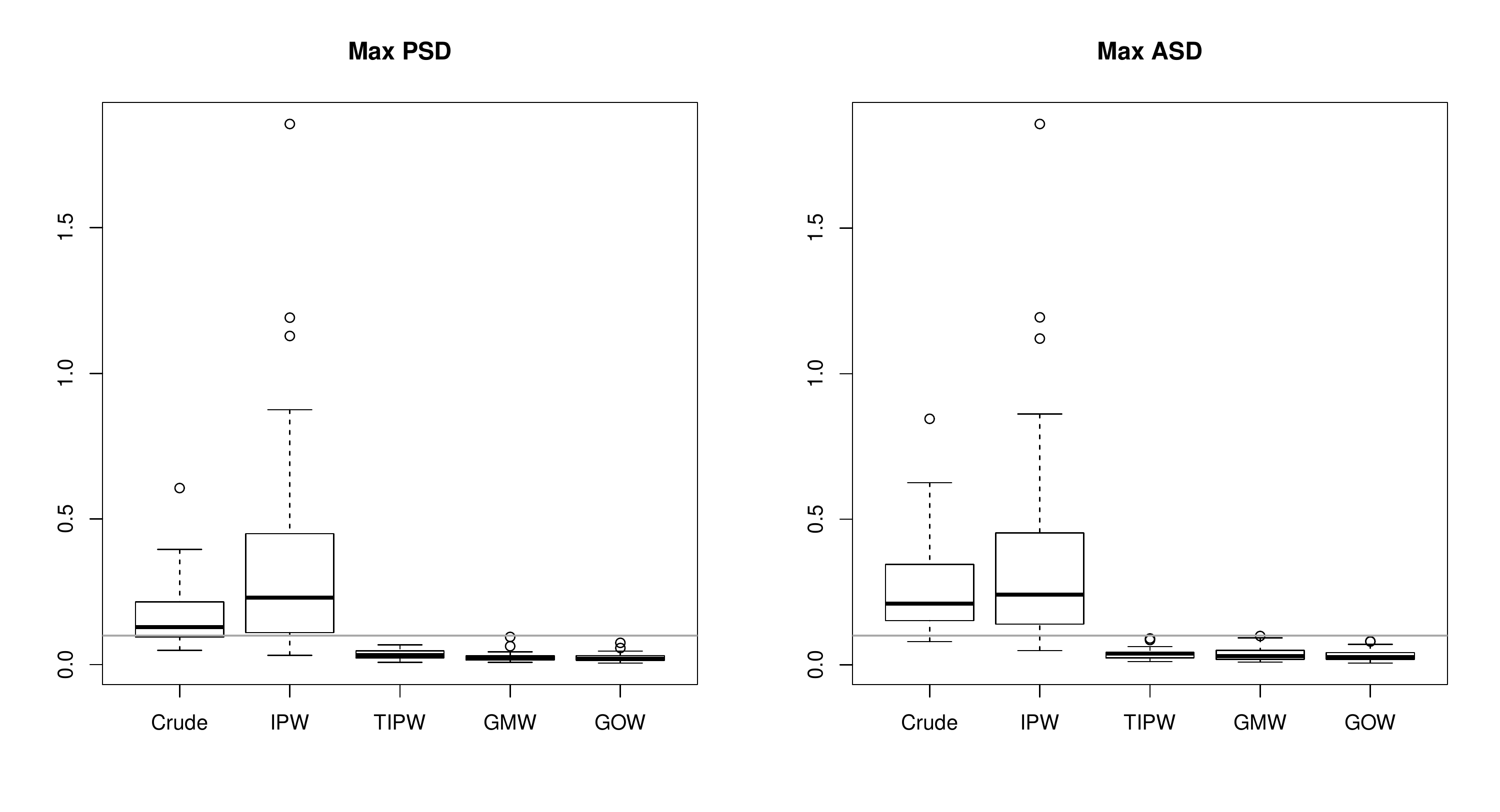}
\caption{Boxplots for the maximum population standardized difference (PSD) and maximum absolute standardized difference (ASD) for all health status covariates corresponding to each adjustment method. The gray horizontal line indicates adequate balance at $0.1$. Crude: unweighted; IPW: inverse probability weighting; TIPW: inverse probability weighting combined with optimal trimming; GMW: generalized matching weighting; GOW: generalized overlap weighting.}
\label{fig:Fig1}
\end{figure}

The MEPS data is well-suited to study racial disparities since it records a wide range of patient-level health characteristics. As previously mentioned, the IOM definition of disparity excludes differences in health status and patient preferences, but includes differences in socioeconomic status and discrimination. For this reason, we follow \citet{McGuire2006} and distinguish between the set of health status variables ($\bX_H$) and the set of SES variables ($\bX_S$), with the former including body mass index, SF-12 physical and mental component summary, comprehensive measurements of health conditions, age, gender, marital status and the latter including poverty status, education, health insurance and geographical region. As there is no gold standard in measuring patient preferences \citep{McGuire2006}, we do not interpret any variables as preference measurements, but acknowledge that the lack of this information represents a limitation in implementing the IOM definition. From the first column of the two boxplots in Figure \ref{fig:Fig1}, we observe substantial differences in the health status distributions among the four racial groups, which indicate the necessity of adjustment.

\subsection{Balance Check and Effective Sample Size}\label{sec:balance}

\begin{figure}[!h]
\centering
\includegraphics[scale=0.5]{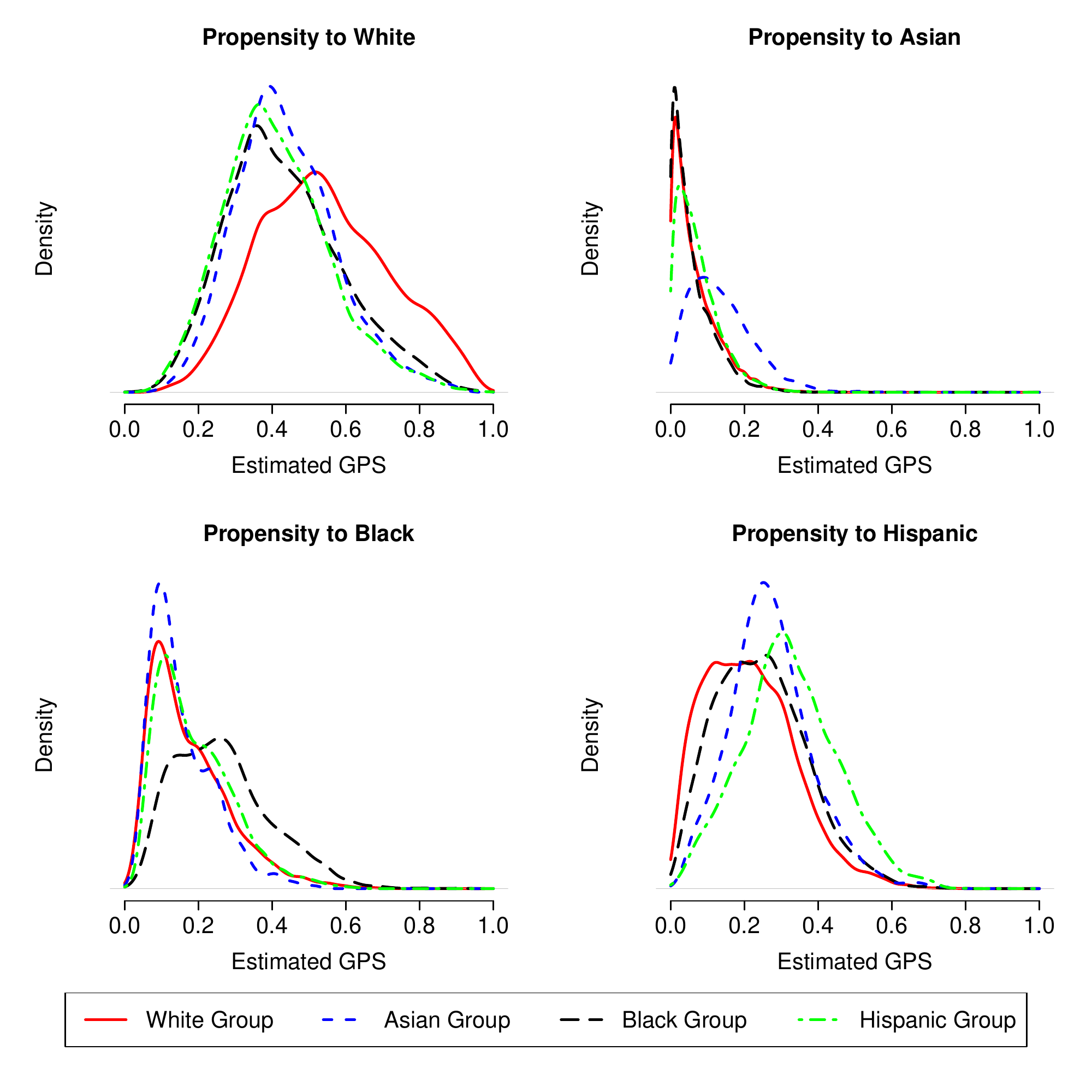}
\caption{Marginal distributions of the estimated health status generalized propensity scores.}
\label{fig:Fig2}
\end{figure}

We employ the generalized propensity scores to balance the health status variables among the four racial groups. If the generalized propensity scores are well estimated, then the propensity-score-weighted populations should be balanced with respect to the health status variables, thus removing the contribution of health status differences to the disparity estimates. This is the general idea behind the application of a health status propensity score to estimate White-minority disparity in the health services literature \citep{Cook2012}. We estimate the generalized propensity scores using a multinomial logistic regression including the main effects of all health status variables. The distributions of the estimated scores are presented in Figure \ref{fig:Fig2}. There is a moderate lack of overlap especially regarding the Asian group. As such, balancing the health status variables toward the combined population through IPW inevitably emphasizes the patients atypical for their own racial groups, producing disparity estimates lacking policy relevance. By contrast, balancing the health status variables toward the overlap population via the generalized overlap weighting (GOW) emphasizes a naturally comparable subpopulation that are most typical in each respective group, and leads to disparity estimates of greater policy interest. Based on the estimated propensity scores, we calculate for each health status variable the values of $\max_j|\text{PSD}_j|$ and $\max_{j<j'}|\text{ASD}_{j,j'}|$, which are defined in Section \ref{sec:BalanceCheck} to examine balance in the weighted populations. Due to the lack of overlap, IPW results in severe imbalances in more than a few health status variables, presenting worse results than no weighting at all. On the other hand, GOW provides the best balance among the overlap population. Two other competing methods, optimal trimming (TIPW) and generalized matching weighting (GMW) also perform adequately in balancing the health status variables in their respective target populations. The balance results are similar between the two balance criteria.

\begin{table}[htbp]
\centering
\caption{Effective sample size of each (weighted) group. Crude: unweighted; IPW: inverse probability weighting; TIPW: inverse probability weighting combined with optimal trimming; GMW: generalized matching weighting; GOW: generalized overlap weighting.}\label{tb:ess}
\begin{tabular}{lrrrrr}
\toprule
 & Whites & Asians & Blacks & Hispanics & Total\\
\midrule
Crude & 9830 & 1446 & 4020 & 5150 & 20446\\
IPW & 8371 & 10 & 2549 & 2482 & 13412\\
TIPW & 6524 & 695 & 2183 & 3071 & 12473\\
GMW & 4937 & 1285 & 1875 & 3176 & 11273\\
GOW & 6015 & 1166 & 2234 & 3756 & 13171\\
\bottomrule
\end{tabular}
\end{table}

To quantify the amount of information in different target populations, we report the corresponding effective sample size (ESS). Following \citet{McCaffrey2013}, we define the ESS for group $j$ as
\bee
\text{ESS}^h_j=\frac{\left(\sum_{i=1}^n\sum_{j=1}^J D_{ij}w_j(\bX_i)\right)^2}
{\sum_{i=1}^n \sum_{j=1}^J D_{ij}w_j^2(\bX_i)}.
\eee
As weighting generally increases the variance compared to the unweighted estimates based on the same sample, the ESS serves as a conservative measure to characterize the variance inflation or precision loss due to weighting. It is evident from Table \ref{tb:ess} that all weighting methods reduce ESS compared to the original sample. However, IPW results in a very small ESS for Asians relative to the original group size, signaling the presence of extreme weights and lack of overlap. By contrast, TIPW, GMW and GOW result in more balanced ESS across groups. Among these alternatives, GOW corresponds to the largest total ESS, matching its theoretical efficiency optimality.

\subsection{Analysis 1: Health Status Propensity Score Weighting}\label{sec:psw}
\begin{table}[htbp]
\centering
\caption{Racial disparity estimates in total health care expenditure (in dollars). The point estimates are obtained as average controlled differences by propensity score weighting. The associated $95\%$ confidence intervals are obtained by the sandwich variance (IPW, TIPW and GOW) or bootstrap (GMW).}\label{tb:dis1}
\scalebox{1}{
\begin{tabular}{lllll}
\toprule
 & IPW & TIPW & GMW & GOW \\
 \midrule
\multirow{2}{*}{Whites-Asians}  & 2402 & 1335 & 1112 & 1160 \\\smallskip
& (530, 4274) & (671, 1999) & (648, 1569) & (660, 1661) \\
\multirow{2}{*}{Whites-Blacks} & 908 & 1148 & 839 & 886 \\\smallskip
 & (505, 1311) & (781, 1515) & (455, 1239) & (518, 1253) \\
\multirow{2}{*}{Whites-Hispanics} & 719 & 1257 & 1234 & 1221 \\\smallskip
& (129, 1309) & (804, 1711) & (813, 1623) & (849, 1593) \\
\multirow{2}{*}{Asians-Blacks} & -1494 & -187 & -273 & -274 \\\smallskip
 & (-3385, 397) & (-872, 499) & (-737, 281) & (-813, 264) \\
\multirow{2}{*}{Asians-Hispanics} & -1683 & -77 & 122 & 61 \\\smallskip
& (-3621, 255) & (-812, 657) & (-385, 621) & (-479, 601) \\
\multirow{2}{*}{Blacks-Hispanics} & -189 & 109 & 395 & 335 \\\smallskip
& (-836, 459) & (-375, 594) & (-100, 820) & (-82, 752) \\
\bottomrule
\end{tabular}}
\end{table}

We calculate the pairwise racial disparities as the weighted average controlled difference in total health care expenditure using GOW, and report point estimates and 95\% confidence intervals (based on the sandwich variance) in the last column of Table \ref{tb:dis1}. This weighting scheme emphasizes a naturally comparable subpopulation with similar health status, namely patients who, based on their health conditions and clinical need, could easily be either White or from each minority group. In other words, this subpopulation features patients whose clinical need variables correspond to the intersection of the White and minority samples' need distributions. Among this overlap subpopulation where all four racial groups have similar health status, Whites spent on average \$1160, \$886 and \$1221 more than Asians, Blacks and Hispanics on health care, with directions and magnitudes comparable to earlier reports from 2003 and 2004 \citep{Cook2009}. All three 95\% confidence intervals exclude zero, confirming that the disparity estimates are significantly different from the null. On the other hand, disparity estimates among the minority groups are not significantly different from zero among the overlap population. For example, the Asians on average spent \$61 more on health care than Hispanics after adjusting for their differences in health status, with zero included in the associated confidence interval.

Disparity estimates may be sensitive to the target population toward which the health status variables are balanced, and notably so with IPW. Here, IPW forces us to balance the health status toward a hypothetical combined population, which is an unrealistic target for policy intervention since it emphasizes patients atypical for their own racial group. The disparity estimates are also likely subject to bias since we found IPW fails to adequately balance the health status variables in Section \ref{sec:balance}. Besides, the lack of overlap leads to loss of efficiency. For example, the largest normalized inverse probability weight is 0.32, accounting for almost one third of the total weights out of 1446 Asians. As a consequence, it is not surprising to for IPW to report the Whites-Asians disparity that is more than twice the magnitude of the GOW estimate. The overlap issue is also apparent when we apply the optimal trimming (\ref{eq:trim}), which excludes about $20\%$ of the sample (2125 Whites, 44 Asians, 1001 Blacks and 603 Hispanics). Unlike IPW, both TIPW and GMW provide disparity estimates closer to GOW, although with wider confidence intervals.

\subsection{Analysis 2: Health Status Propensity Score Weighting with Rank-and-Replace Adjustment} \label{sec:psw_rr}
While the health propensity score weighting in Section \ref{sec:psw} allows us to balance health status variables without peeking at the outcome distribution, it does not account for the contribution of SES variables. The IOM definition requires adjustment for $\bX_H$ but includes justifiable differences in the distributions of SES variables $\bX_S$; the latter reflect differential impact of operations of health care systems and regulatory climate \citep{IOM2003}. If variables in $\bX_H$ are independent of variables in $\bX_S$, then the analysis in Section \ref{sec:psw} is IOM-concordant; if the variables in $\bX_H$ are correlated with variables in $\bX_S$, health status propensity score weighting may inadvertently alter the distributions of $\bX_S$ and only provides an approximation to the IOM-defined disparity \citep{Balsa2007}. To address such a concern, we apply the rank-and-replace adjustment method \citep{McGuire2006} to undo the undesired weighting of $\bX_S$ by the health status propensity score. \citet{Cook2010} combined binary overlap weights with rank-and-replace SES adjustment; here we extend the method to comparing multiple racial groups.

\begin{table}[htbp]
\centering
\caption{Racial disparity estimates in total health care expenditure (in dollars). The point estimates are obtained as weighted average controlled differences by the combined propensity score and rank-and-replace method. The associated $95\%$ confidence intervals are obtained by bootstrap.}\label{tb:dis2}
\scalebox{1}{
\begin{tabular}{lllll}
\toprule
 & IPW & TIPW & GMW & GOW \\
 \midrule
\multirow{2}{*}{Whites-Asians}  & -1194 & 1133 & 997 & 1023 \\\smallskip
& (-5307, 2534) & (258, 1877) & (486, 1530) & (464, 1584) \\
\multirow{2}{*}{Whites-Blacks} & 1610 & 1610 & 1013 & 1069 \\\smallskip
 & (1184, 1980) & (1248, 1942) & (668, 1299) & (728, 1357) \\
\multirow{2}{*}{Whites-Hispanics} & 1899 & 1883 & 1374 & 1420 \\\smallskip
& (1381, 2352) & (1446, 2232) & (1082, 1673) & (1128, 1731) \\
\multirow{2}{*}{Asians-Blacks} & 2804 & 476 & 16 & 46 \\\smallskip
 & (-965, 6926) & (-367, 1323) & (-578, 551) & (-582, 594) \\
\multirow{2}{*}{Asians-Hispanics} & 3093 & 749 & 377 & 397 \\\smallskip
& (-689, 7149) & (-83, 1565) & (-184, 902) & (-206, 967) \\
\multirow{2}{*}{Blacks-Hispanics} & 289 & 273 & 361 & 351 \\\smallskip
& (-273, 805) & (-177, 629) & (41, 722) & (27, 721) \\
\bottomrule
\end{tabular}}
\end{table}

Following \citet{Cook2009}, we perform the rank-and-replace adjustment based on a model-based SES index to equalize the weighted SES distributions and the unweighted marginals. We model the health care expenditure as a function of $\bX_H$, $\bX_S$ and racial group indicator: $g(\bE[Y_i|\bX_{H,i},\bX_{S,i},Z_i])=\gamma_0+\bX_{H,i}^T\bgamma_H+\bX_{S,i}^T\bgamma_S+\sum_{j=1}^J\gamma_{1j}D_{ij}$, where the SES predictive index is denoted by $\bX_{S,i}^T\bgamma_S$. We choose $g$ as the log link, and to allow for heteroscedastic variances \citep{Buntin2004}, apply the Park test to determine the variance power relative to the mean \citep{Park1966,Manning2001}. In other words, the model parameters are estimated by a Tweedie generalized linear model with data-driven specification of the power variance function \citep{Jorgensen2015}. The estimated coefficients provide the SES index value for each patient, and we obtain the weighted rank of $\bX_{S,i}^T\bgamma_S$ within each racial group. The rank-and-replace method then restores the original group-specific SES distributions by replacing the propensity score weighted SES index values with the equivalently ranked unweighted SES index values. With this adjustment, the weighted distribution of the SES index values in each group is approximately the same as the original distribution of the index values in that group, and the resulting disparity estimates become IOM-concordant by recapturing the racial differences in SES.

We obtain the SES-adjusted expected expenditure for each patient through the generalized linear model, and calculate the weighted average controlled differences based on the adjusted expenditure. After balancing the health status variables toward the overlap population, factoring the SES differences into the calculation increases the Whites-Blacks, Whites-Hispanics disparity by \$183 and \$199 and decreases the Whites-Asians disparity by \$137, without modifying the direction and statistical significance. Such changes may be anticipated, for example, between Whites and Blacks in the following case. Given Whites have overall higher health status and SES and that $\bX_H$, $\bX_S$ are likely positively correlated, White patient with lower health status and lower SES will be weighted more heavily to balance $\bX_H$. Assuming that White patients with lower SES have lower health care utilization, we would expect the slight increase in the Whites-Blacks disparity after restoring the original SES distributions. On the other hand, the SES adjustment had a larger effect on disparities among the minority groups, but the results remain statistically insignificant. Overall, the changes in the GOW estimates from Table \ref{tb:dis1} and Table \ref{tb:dis2} suggest that racial differences in health care utilization were slightly mediated through the SES variables. The interpretations of the disparity estimates are similar to those in Section \ref{sec:psw}, except that differences due to SES variables contribute to the disparity measures by the IOM definition.

In contrast to the results obtained by the generalized overlap weights, the SES adjustment magnifies the undue influence of extreme propensities when IPW is used to balance $\bX_H$, since for example, Whites are found to on average spend \$1194 less than Asians among the combined population. With IPW, not only the hypothetical combined population is of minimal policy relevance, but also the inherent bias due to extreme propensities complicates the interpretation of the unusual direction in such a point estimate.

\section{Simulations}\label{sec:sim}
To further shed light on the comparison between different weighting methods, we conduct simulations in the context of observational studies with multiple non-randomized treatments. Our data generating process is similar to \citet{Yang2016} except that we consider nonzero pairwise average treatment effect among the considered target populations. We generate covariates $X_{i1}$, $X_{2i}$ and $X_{3i}$ from a multivariate normal distribution with mean vector $(2,1,1)$ and covariances of $(1,-1,-0.5)$; $X_{4i}\sim\text{Uniform}[-3,3]$; $X_{5i}\sim\chi_1^2$ and $X_{6i}\sim \text{Bernoulli}(0.5)$, with the covariate vector $\bX_i^T=(X_{1i},X_{2i},X_{3i},X_{4i},X_{5i},X_{6i})$. The assignment mechanism follows the multinomial logistic model
\bee
(D_{i1},\ldots,D_{iJ})|\bX_i\sim\text{Multinom}(e_1(\bX_i),\ldots,e_J(\bX_i)),
\eee
where $D_{ij}$ is the treatment indicator defined in Section \ref{sec:setup} and $e_j(\bX_i)=\exp(\alpha_j+\bX_i^T\bbeta_j)/\sum_{k=1}^J\exp(\alpha_k+\bX_i^T\bbeta_k)$ is the true generalized propensity score with $\alpha_1=0$, $\bbeta_1^T=(0,0,0,0,0,0)$. In the first simulation with $J=3$ treatment groups, $\bbeta_2^T=\kappa_2\times (1,1,1,-1,-1,1)$ and $\bbeta_3^T=\kappa_3\times (1,1,1,1,1,1)$. We set $(\kappa_2,\kappa_3)=(0.2,0.1)$ to simulate a scenario with adequate covariate overlap and $(\kappa_2,\kappa_3)=(0.8,0.4)$ to induce lack of overlap with strong propensity tails, i.e., the propensity to receive certain treatment is close to zero for specific design values. We further choose $\alpha_2$ and $\alpha_3$ so that the overall treatment proportions are fixed at $(0.3, 0.4, 0.3)$. The potential outcomes are generated from $Y_i(j)=\left(1,\bX_i^T\right)\bgamma_j+\epsilon_i$ with $\epsilon_i\sim N(0,1)$, $\bgamma_1^T=(-1.5,1,1,1,1,1,1)$, $\bgamma_2^T=(-4,2,3,1,2,2,2)$ and $\bgamma_3^T=(3,3,1,2,-1,-1,-1)$. In the second simulation with $J=6$ groups, we similarly specify the parameters to simulate both adequate and lack of overlap. The detailed specification and visual inspection of the overlap in each simulation scenario can be found in Section D of the Supplementary Material \citep{LiLiSupp2019}. The total sample size is fixed at $n=1500$ for $J=3$ and $n=6000$ for $J=6$.

\begin{table}[h!]
\caption{Simulation results with $J=3$ treatment groups. With adequate overlap, the optimally trimming excludes at most $2\%$ of the total sample. Under lack of overlap, the optimal trimming rule excludes $19\%$ to $30\%$ of the total sample.}
\centering\label{tb:sim}
\begin{tabular}{llcccccc}
\toprule
Metric & Method & \multicolumn{3}{c}{Adequate Overlap} & \multicolumn{3}{c}{Lack of Overlap}\\
\cmidrule(lr){3-5}\cmidrule(lr){6-8}
& & $\tau(\blambda_{1,2})$ & $\tau(\blambda_{1,3})$ & $\tau(\blambda_{2,3})$ & $\tau(\blambda_{1,2})$ & $\tau(\blambda_{1,3})$ & $\tau(\blambda_{2,3})$ \\
\midrule
\multirow{7}{*}{$|\text{Bias}|$}
& DIF & 0.46 & 0.60 & 0.14 & 0.43 & 0.64 & 0.21 \\
& IPW & 0.02 & 0.01 & 0.01 & 0.19 & 0.02 & 0.17 \\
& TIPW & 0.01 & 0.002 & 0.01 & 0.03 & 0.01 & 0.01 \\
& GPSM & 0.02 & 0.01 & 0.01 & 0.25 & 0.10 & 0.15 \\
& TGPSM & 0.02 & 0.004 & 0.01 & 0.08 & 0.02 & 0.05 \\
& GMW & 0.02 & 0.01 & 0.02 & 0.001 & 0.01 & 0.01 \\\smallskip
& GOW & 0.01 & 0.001 & 0.01 & 0.01 & 0.01 & 0.003 \\
\multirow{7}{*}{RMSE}
& DIF & 0.55 & 0.65 & 0.37 & 0.50 & 0.68 & 0.38 \\
& IPW & 0.20 & 0.16 & 0.26 & 1.04 & 0.61 & 1.16\\
& TIPW & 0.16 & 0.16 & 0.23 & 0.38 & 0.28 & 0.47\\
& GPSM & 0.26 & 0.22 & 0.31 & 0.86 & 0.51 & 0.90\\
& TGPSM & 0.25 & 0.23 & 0.31 & 0.53 & 0.37 & 0.60\\
& GMW & 0.17 & 0.18 & 0.27 & 0.29 & 0.24 & 0.36\\\smallskip
& GOW & 0.15 & 0.15 & 0.22 & 0.28 & 0.23 & 0.35\\
\multirow{7}{*}{Coverage}
& DIF & 0.64 & 0.36 & 0.92 & 0.65 & 0.23 & 0.90 \\
& IPW & 0.92 & 0.95 & 0.95 & 0.79 & 0.88 & 0.91 \\
& TIPW & 0.94 & 0.94 & 0.94 & 0.93 & 0.90 & 0.91\\
& GPSM & 0.99 & 0.97 & 0.97 & 0.88 & 0.91 & 0.91 \\
& TGPSM & 0.98 & 0.96 & 0.98 & 0.95 & 0.92 & 0.95 \\
& GMW & 0.95 & 0.96 & 0.94 & 0.95 & 0.95 & 0.95 \\
& GOW & 0.94 & 0.96 & 0.95 & 0.95 & 0.94 & 0.94 \\
\bottomrule
\end{tabular}
\end{table}

For each scenario, we simulate $1000$ datasets and estimate the pairwise causal effects using alternative estimators. To quantify the confounding bias in each simulation scenario, we first report the raw difference in means (DIF). For comparison among weighting methods, we consider GOW, IPW, TIPW and GMW. We also examine a recent propensity score matching estimator proposed by \citet{Yang2016}, both without and with the optimal trimming step (GPSM and TGPSM). GPSM separately exploits each scalar propensity score for estimating the average potential outcomes and thus resolves the issue of matching on high-dimensional propensity score vector. Because the target population may differ in different estimators, we assess the accuracy of estimators relative to their corresponding target estimands. Specifically, the target estimands of DIF, IPW and GPSM are pairwise ATE for the combined population and are analytically determined from the true potential outcome model, whereas the target estimands for GMW, GOW, TIPW and TGPSM are defined for subpopulations and evaluated numerically based on Monte Carlo integration. For each data replicate, we estimate the generalized propensity scores based on the correct multinomial logistic regression model including all covariates. The proposed sandwich variance \eqref{eq:sand} was used to obtain confidence intervals for GOW. The empirical sandwich variance (see Section C of the Supplementary Material \citep{LiLiSupp2019} for details) and the \citet{Abadie2012} variance were used to obtain interval estimators for IPW and GPSM. Since the weight function $w_j(\bX)$ for GMW is not everywhere differentiable (with infinite-many non-differentiable points) and fails to satisfy the regularity conditions for deriving a sandwich variance, we use bootstrap for interval estimation. Finally, whenever trimming is used, the generalized propensity scores are re-estimated based on the trimmed sample as refitting improves the finite-sample performance of the resulting estimators \citep{LiThomasLi2018}; accordingly, variance calculation is carried out based on the trimmed sample.

Table \ref{tb:sim} summarizes the absolute bias, root mean squared error (RMSE) and coverage of each estimator with $J=3$ groups. As expected, DIF shows substantial bias and under-coverage, indirectly characterizing the magnitude of confounding bias. All other approaches perform reasonably well when there is adequate overlap. With lack of overlap, IPW and GPSM are sensitive to extreme propensities and produce biased point estimates. The optimal trimming method excludes $19\%$ to $30\%$ of the total sample, reduces the bias and improves efficiency and coverage in estimating the subpopulation causal effects. By down-weighting extreme units, both GMW and GOW provide unbiased point estimates with nominal coverage. Overall, TIPW, GMW and GOW are associated with the smallest RMSE and are more efficient than the other methods. Among them, GOW has the smallest RMSE, matching the theoretical predictions in Section \ref{sec:asymp}.

The simulation results with $J=6$ groups are presented in Web Figures 5 and 6 in Section D of the Supplementary Material \citep{LiLiSupp2019}. With adequate overlap, all methods have good control of confounding bias, produce unbiased estimates and close to nominal coverage. GMW and GOW provide the lowest RMSE, with the latter demonstrating higher efficiency for estimating most of the causal contrasts (the ratio of total MSE is $1.18$). With lack of overlap, the clear separation of covariate space makes it challenging to simultaneously remove all confounding for estimating the $15$ pairwise contrasts. By discarding more than half of the sample, the optimal trimming method improves the bias, efficiency and coverage properties over IPW and GPSM, both of which are subject to bias and excessive variance with extreme propensities. GMW and GOW further improve the efficiency and coverage properties upon trimming by down-weighting the extreme units. Concordant with the large-sample theory, GOW produces more efficient estimates than GMW for $12$ out of $15$ causal contrasts (the ratio of total MSE is $1.17$). In this challenging scenario, the bootstrap CI for GMW has slightly better finite-sample coverage than the closed-form CI for GOW based on the empirical sandwich variance, but the closed-form CI estimator for GOW demonstrates the best coverage among all the considered closed-form CI estimators. However, another substantial gain of GOW over GMW is the computational time: for each simulation, the bootstrap interval estimates for GMW with 1000 samples require more than $80$ times longer running time than that of the closed-form GOW interval estimates, which can be very burdensome for large observational datasets.

\section{Discussion}\label{sec:dis}
We proposed a unified propensity score weighting framework, the balancing weights, for causal inference with multiple treatments. Within this framework, we developed the generalized overlap weights for pairwise comparisons to emphasize the target population with the most covariate overlap. We applied these new weights to study health care disparities and found Whites had significantly more spendings on health care than the minority groups in 2009, after adjusting for differential distributions of health status. In contrast, the disparity estimates are not significantly different from zero between the minorities. This patten persists regardless of considerations of the SES differences. These results could potentially help health policy decision makers direct more resources and infrastructures for the minority groups to improve their access to medical care as a means to minimize the White-minority disparities in utilization.

Following the conceptual framework introduced in \citet{McGuire2006}, the interpretation of the health care disparity estimates in this application remains descriptive. Typically, health care disparity includes justifiable differences due to operation of health care systems and regulatory climate (often measured by SES) and discrimination (residual inequality) but excludes differences in clinical appropriateness and need (measured by health status variables). By this definition, we aim to quantify how much the average spending differs between racial groups vis-\`{a}-vis a common reference population with the same clinical need. This objective motivates the propensity score weighting methodology, which is a popular adjustment tool in comparative effectiveness research. Because the disparity estimates are calculated based on that common reference population, it is critical to conceptualize different populations implied by different weighting schemes. The IPW creates a combined population from all racial groups where the resulting patients has need variables corresponding to the union of the White and minority samples' need distributions. This union population inevitably features patients in other racial groups and hence may not be representative within each racial group. To improve upon IPW which targets this unrealistic population, we developed the generalized overlap weights to target a subpopulation with health status corresponding to the intersection of the White and minority samples' health status distributions. As this overlap subpopulation remains representative for each racial group, it could be regarded an actionable subset to track health care disparity. To further produce IOM-concordant disparity estimates, we combined the rank-and-replacement adjustment with propensity score weighting to describe the average differences in health care utilization after adjusting for clinical need but restoring the SES differences in Section \ref{sec:psw_rr}.

We do not intend to make a causal statement of the racial disparity in health care utilization, but there may be a tendency to do so based on the parallel discussion on health disparity or inequality. While one should generally distinguish between \emph{health care} disparity and \emph{health} disparity as the corresponding methodologies differ \citep{McGuire2006}, it is possible to borrow the weak causal perspective of \citet{VanderWeele2014a,VanderWeele2014b} developed around health inequality to interpret the health care disparity in Section \ref{sec:app}. For instance, the estimates in Table \ref{tb:dis1} could be understood as the remaining differences in health care utilization if we were to, hypothetically, intervene on the differential health status across groups. Because such an interpretation is not typical in studying health care disparity, we keep the descriptive interpretation as in \citet{McGuire2006,Cook2010} and \citet{LiZaslavskyLandrum2013}.

Even though our application responds to challenges in describing patterns for health care utilization, the proposed propensity score methods are highly relevant in comparative effectiveness research based on observational data. For example, the target estimand---the pairwise ATO---describes the causal comparison in the subpopulation with clinical equipoise, and may be preferred \citep{LiThomasLi2018}. With the increasing use of convenience samples in observational studies, the proposed generalized overlap weights represent a flexible adjustment method to regain a target population where current practice remains uncertain, rather than a target population dominated by extreme units for whom treatment decisions are already clear. Our presentation has focused exclusively on categorical treatments but the concept of target population remains relevant with a continuous treatment. In the latter setting, the weighted estimands (\ref{eq:meanpo}) may also be cast as the average potential outcomes among the combined population under a stochastic intervention or modified treatment policy \citep{Munoz2012,Haneuse2013}, which could provide an alternative interpretation.

There are several directions for extending the proposed method. First, as with all propensity score methods, a well-estimated propensity score is crucial to the analysis. To focus on the main message, this paper adopted a convenient parametric model to estimate the generalized propensity scores. A natural extension is to use flexible machine learning models to estimate the generalized propensity scores; examples include the Generalized Boosting Model \citep{McCaffrey2004,McCaffrey2013}, ensemble learning methods such as the Super Learner \citep{Dudoit2005,Pirracchio2015}, the debiased machine learning estimator \citep{Chernozhukov2018}, as well as Bayesian nonparametric models.

Second, the generalized overlap weights are obtained by setting the linear contrast coefficients $\ba$ to allow for pairwise comparisons, which are of general scientific interest with multiple categorical treatments. When there is no strong \emph{a priori} preference for $\ba$, one possibility is to choose $\ba$ based on minimizing a specific loss function \citep{Hirshberg2017}.

Third, this paper focused on the moment weighting estimators; these estimators are not semiparametric efficient even with a correct propensity score model \citep{Hirano2003}. An important avenue for improvement is to consider the class of augmented weighting estimators with balancing weights \citep{Robins1994}. One could construct, for each choice of the balancing weights, an augmented estimator as
\bee
\hat{m}^{h,\text{aug}}_j=\hat{m}^h_j-\frac{\sum_{i=1}^n(D_{ij}-e_j(\bX_i))w_j(\bX_i)\hat{m}_j(\bX_i)}{\sum_{i=1}^n h(\bX_i)},
\eee
where $\hat{m}_j(\bX_i)=\hat{\bE}[Y(j)|\bX]$ is the outcome regression function. It can be shown that $\hat{m}^{h,\text{aug}}_j$ is semiparametric efficient for estimating $m_j^h$ when both the generalized propensity score model and the regression function are correctly specified. Of note, when the tilting function $h(\bX_i)=1$, $\hat{m}^{h,\text{aug}}_j$ has an additional doubly-robustness property such that it is consistent to $\bE[Y(j)]$ when either the generalized propensity score model or the regression function is correctly specified, but not necessarily both. However, this robustness property does not generally hold for $\hat{m}^{h,\text{aug}}_j$ when $h$ is a function of the propensity scores, such as the optimal tilting function considered in Section \ref{sec:gow}. In this case, the consistency necessitates a correct propensity score model regardless of the outcome model (also see \citet{LiLi2019} for an example with ATT). Nevertheless, outcome regression may still increase the efficiency of the weighting estimator. For this reason, it would be valuable in future work to explore the application of the augmented weighting estimator to the racial disparity study. For example, in each racial group, we could fit an additional regression model for the health care expenditure as a function of $\bX_{H}$, and estimate pairwise disparity by $\hat{\tau}^h(\blambda_{j,j'})=\hat{m}^{h,\text{aug}}_j-\hat{m}^{h,\text{aug}}_{j'}$ for the analysis in Section \ref{sec:psw}. It is currently unclear how to combine the rank-and-replace adjustment with the augmented weighting approach for the analysis in Section \ref{sec:psw_rr}, since the rank-and-replace adjustment already involves an outcome model.

Finally, the balancing weights framework pursues weighting by propensity scores to achieve balance, with different choices of weights targeting specific populations and causal estimands. An alternative strand of recent literature derives weights that directly balance the covariates, bypassing the estimation of propensity scores; examples include the entropy balancing \citep{Hainmueller2012}, the stabilized balancing weights \citep{Zubizarreta2015} and the approximate residual balancing \citep{Athey2018}. Those weights usually focus on the ATE or ATT estimand with binary treatments, and do not involve adaptively changing the target population as our general balancing weights framework. In practice, it is prudent for the analyst to choose a method according to the scientific question and settings of specific applications rather than fixating on one single method.

\section*{Acknowledgements}
We thank Benjamin Le Cook for providing the MEPS dataset, and Alan Zaslavsky, Laine Thomas, Peng Ding for insightful discussions. The first author is grateful to the ASA Biometrics Section for receiving a JSM Student Paper Award based on an earlier version of this article. We thank the Editor, Associate Editor and two anonymous referees for their constructive comments, which have greatly improved the exposition of this work.

\begin{supplement}
\stitle{Supplement to ``Propensity score weighting for causal inference with multiple treatments"}
\slink[doi]{COMPLETED BY THE TYPESETTER}
\sdatatype{.pdf}
\sdescription{\\
\textbf{Supplement A:} On Transitivity. We provide a detailed discussion on transitivity of the target estimands for pairwise comparisons.\\
\textbf{Supplement B:} Proof of Propositions. We present detailed proofs of Propositions \ref{prop1} to \ref{prop:hopt} in Section \ref{sec:asymp}.\\
\textbf{Supplement C:} Proof of Theorem \ref{thm:var}. We provide the derivation and related discussions of the variance estimator for the generalized overlap weighting.\\
\textbf{Supplement D:} Additional Simulation Results. We present additional figures and numerical results for the simulation study in Section \ref{sec:sim}.}
\end{supplement}

\bibliographystyle{imsart-nameyear}
\bibliography{MultiTreatment}

\begin{thebibliography}{49}

\bibitem[\protect\citeauthoryear{Abadie and Imbens}{2012}]{Abadie2012}
\begin{barticle}[author]
\bauthor{\bsnm{Abadie},~\bfnm{Alberto}\binits{A.}} \AND
  \bauthor{\bsnm{Imbens},~\bfnm{Guido~W.}\binits{G.~W.}}
(\byear{2012}).
\btitle{{A martingale representation for matching estimators}}.
\bjournal{Journal of the American Statistical Association}
\bvolume{107}
\bpages{833--843}.
\end{barticle}
\endbibitem

\bibitem[\protect\citeauthoryear{Athey, Imbens and Wager}{2018}]{Athey2018}
\begin{barticle}[author]
\bauthor{\bsnm{Athey},~\bfnm{Susan}\binits{S.}},
  \bauthor{\bsnm{Imbens},~\bfnm{Guido~W.}\binits{G.~W.}} \AND
  \bauthor{\bsnm{Wager},~\bfnm{Stefan}\binits{S.}}
(\byear{2018}).
\btitle{{Approximate residual balancing: debiased inference of average
  treatment effects in high dimensions}}.
\bjournal{Journal of the Royal Statistical Society. Series B: Statistical
  Methodology}
\bvolume{80}
\bpages{597--623}.
\end{barticle}
\endbibitem

\bibitem[\protect\citeauthoryear{Balsa, Cao and McGuire}{2007}]{Balsa2007}
\begin{barticle}[author]
\bauthor{\bsnm{Balsa},~\bfnm{A.~I.}\binits{A.~I.}},
  \bauthor{\bsnm{Cao},~\bfnm{Z.}\binits{Z.}} \AND
  \bauthor{\bsnm{McGuire},~\bfnm{T.~G.}\binits{T.~G.}}
(\byear{2007}).
\btitle{{Does managed health care reduce health care disparities between
  minorities and Whites?}}
\bjournal{Journal of Health Economics}
\bvolume{27}
\bpages{781--807}.
\end{barticle}
\endbibitem

\bibitem[\protect\citeauthoryear{Buntin and Zaslavsky}{2004}]{Buntin2004}
\begin{barticle}[author]
\bauthor{\bsnm{Buntin},~\bfnm{Melinda~Beeuwkes}\binits{M.~B.}} \AND
  \bauthor{\bsnm{Zaslavsky},~\bfnm{Alan~M.}\binits{A.~M.}}
(\byear{2004}).
\btitle{{Too much ado about two-part models and transformation? Comparing
  methods of modeling Medicare expenditures}}.
\bjournal{Journal of Health Economics}
\bvolume{23}
\bpages{525--542}.
\end{barticle}
\endbibitem

\bibitem[\protect\citeauthoryear{Chernozhukov et~al.}{2018}]{Chernozhukov2018}
\begin{barticle}[author]
\bauthor{\bsnm{Chernozhukov},~\bfnm{Victor}\binits{V.}},
  \bauthor{\bsnm{Chetverikov},~\bfnm{Denis}\binits{D.}},
  \bauthor{\bsnm{Demirer},~\bfnm{Mert}\binits{M.}},
  \bauthor{\bsnm{Duflo},~\bfnm{Esther}\binits{E.}},
  \bauthor{\bsnm{Hansen},~\bfnm{Christian}\binits{C.}},
  \bauthor{\bsnm{Newey},~\bfnm{Whitney}\binits{W.}} \AND
  \bauthor{\bsnm{Robins},~\bfnm{James}\binits{J.}}
(\byear{2018}).
\btitle{{Double/debiased machine learning for treatment and structural
  parameters}}.
\bjournal{Econometrics Journal}
\bvolume{21}
\bpages{1--68}.
\end{barticle}
\endbibitem

\bibitem[\protect\citeauthoryear{Cook, McGuire and Zaslavsky}{2012}]{Cook2012}
\begin{barticle}[author]
\bauthor{\bsnm{Cook},~\bfnm{Benjamin~L.}\binits{B.~L.}},
  \bauthor{\bsnm{McGuire},~\bfnm{Thomas~G.}\binits{T.~G.}} \AND
  \bauthor{\bsnm{Zaslavsky},~\bfnm{Alan~M.}\binits{A.~M.}}
(\byear{2012}).
\btitle{{Measuring racial/ethnic disparities in health care: Methods and
  practical issues}}.
\bjournal{Health Services Research}
\bvolume{47}
\bpages{1232--1254}.
\end{barticle}
\endbibitem

\bibitem[\protect\citeauthoryear{Cook et~al.}{2009}]{Cook2009}
\begin{barticle}[author]
\bauthor{\bsnm{Cook},~\bfnm{Benjamin~L.}\binits{B.~L.}},
  \bauthor{\bsnm{Mcguire},~\bfnm{Thomas~G}\binits{T.~G.}},
  \bauthor{\bsnm{Meara},~\bfnm{Ellen}\binits{E.}} \AND
  \bauthor{\bsnm{Zaslavsky},~\bfnm{Alan~M}\binits{A.~M.}}
(\byear{2009}).
\btitle{{Adjusting for health status in non-linear models of health care
  disparities}}.
\bjournal{Health Services and Outcomes Research Methodology}
\bvolume{9}
\bpages{1--21}.
\end{barticle}
\endbibitem

\bibitem[\protect\citeauthoryear{Cook et~al.}{2010}]{Cook2010}
\begin{barticle}[author]
\bauthor{\bsnm{Cook},~\bfnm{Benjamin~L.}\binits{B.~L.}},
  \bauthor{\bsnm{Mcguire},~\bfnm{Thomas~G}\binits{T.~G.}},
  \bauthor{\bsnm{Lock},~\bfnm{Kari}\binits{K.}} \AND
  \bauthor{\bsnm{Zaslavsky},~\bfnm{Alan~M}\binits{A.~M.}}
(\byear{2010}).
\btitle{{Comparing methods of racial and ethnic disparities measurement across
  different settings of mental health care}}.
\bjournal{Health Services Research}
\bvolume{45}
\bpages{825--847}.
\end{barticle}
\endbibitem

\bibitem[\protect\citeauthoryear{Crump et~al.}{2009}]{Crump2009}
\begin{barticle}[author]
\bauthor{\bsnm{Crump},~\bfnm{Richard~K.}\binits{R.~K.}},
  \bauthor{\bsnm{Hotz},~\bfnm{V.~Joseph}\binits{V.~J.}},
  \bauthor{\bsnm{Imbens},~\bfnm{Guido~W.}\binits{G.~W.}} \AND
  \bauthor{\bsnm{Mitnik},~\bfnm{Oscar~A.}\binits{O.~A.}}
(\byear{2009}).
\btitle{{Dealing with limited overlap in estimation of average treatment
  effects}}.
\bjournal{Biometrika}
\bvolume{96}
\bpages{187--199}.
\end{barticle}
\endbibitem

\bibitem[\protect\citeauthoryear{Ding and Li}{2018}]{ding2018causal}
\begin{barticle}[author]
\bauthor{\bsnm{Ding},~\bfnm{Peng}\binits{P.}} \AND
  \bauthor{\bsnm{Li},~\bfnm{Fan}\binits{F.}}
(\byear{2018}).
\btitle{Causal inference: A missing data perspective}.
\bjournal{Statistical Science}
\bvolume{33}
\bpages{214--237}.
\end{barticle}
\endbibitem

\bibitem[\protect\citeauthoryear{Dudoit and van~der Laan}{2005}]{Dudoit2005}
\begin{barticle}[author]
\bauthor{\bsnm{Dudoit},~\bfnm{Sandrine}\binits{S.}} \AND
  \bauthor{\bparticle{van~der} \bsnm{Laan},~\bfnm{Mark~J.}\binits{M.~J.}}
(\byear{2005}).
\btitle{{Asymptotics of cross-validated risk estimation in estimator selection
  and performance assessment}}.
\bjournal{Statistical Methodology}
\bvolume{2}
\bpages{131--154}.
\end{barticle}
\endbibitem

\bibitem[\protect\citeauthoryear{Feng et~al.}{2012}]{Feng2012}
\begin{barticle}[author]
\bauthor{\bsnm{Feng},~\bfnm{Ping}\binits{P.}},
  \bauthor{\bsnm{Zhou},~\bfnm{Xiao~Hua}\binits{X.~H.}},
  \bauthor{\bsnm{Zou},~\bfnm{Qing~Ming}\binits{Q.~M.}},
  \bauthor{\bsnm{Fan},~\bfnm{Ming~Yu}\binits{M.~Y.}} \AND
  \bauthor{\bsnm{Li},~\bfnm{Xiao~Song}\binits{X.~S.}}
(\byear{2012}).
\btitle{{Generalized propensity score for estimating the average treatment
  effect of multiple treatments}}.
\bjournal{Statistics in Medicine}
\bvolume{31}
\bpages{681--697}.
\end{barticle}
\endbibitem

\bibitem[\protect\citeauthoryear{Hainmueller}{2012}]{Hainmueller2012}
\begin{barticle}[author]
\bauthor{\bsnm{Hainmueller},~\bfnm{J.}\binits{J.}}
(\byear{2012}).
\btitle{{Entropy balancing for causal effects: A multivariate reweighting
  method to produce balanced samples in observational studies.}}
\bjournal{Political Analysis}
\bvolume{1}
\bpages{25--46}.
\end{barticle}
\endbibitem

\bibitem[\protect\citeauthoryear{Haneuse and Rotnitzky}{2013}]{Haneuse2013}
\begin{barticle}[author]
\bauthor{\bsnm{Haneuse},~\bfnm{S.}\binits{S.}} \AND
  \bauthor{\bsnm{Rotnitzky},~\bfnm{A.}\binits{A.}}
(\byear{2013}).
\btitle{{Estimation of the effect of interventions that modify the received
  treatment}}.
\bjournal{Statistics in Medicine}
\bvolume{32}
\bpages{5260--5277}.
\end{barticle}
\endbibitem

\bibitem[\protect\citeauthoryear{Hirano, Imbens and Ridder}{2003}]{Hirano2003}
\begin{barticle}[author]
\bauthor{\bsnm{Hirano},~\bfnm{K}\binits{K.}},
  \bauthor{\bsnm{Imbens},~\bfnm{GW}\binits{G.}} \AND
  \bauthor{\bsnm{Ridder},~\bfnm{G}\binits{G.}}
(\byear{2003}).
\btitle{Efficient estimation of average treatment effects using the estimated
  propensity score}.
\bjournal{Econometrica}
\bvolume{71}
\bpages{1161--1189}.
\end{barticle}
\endbibitem

\bibitem[\protect\citeauthoryear{Hirshberg and
  Zubizarreta}{2017}]{Hirshberg2017}
\begin{barticle}[author]
\bauthor{\bsnm{Hirshberg},~\bfnm{David~A.}\binits{D.~A.}} \AND
  \bauthor{\bsnm{Zubizarreta},~\bfnm{Jos{\'{e}}~R.}\binits{J.~R.}}
(\byear{2017}).
\btitle{{On two approaches to weighting in causal inference}}.
\bjournal{Epidemiology}
\bvolume{28}
\bpages{812--816}.
\end{barticle}
\endbibitem

\bibitem[\protect\citeauthoryear{Imbens}{2000}]{Imbens2000}
\begin{barticle}[author]
\bauthor{\bsnm{Imbens},~\bfnm{Guido~W.}\binits{G.~W.}}
(\byear{2000}).
\btitle{{The role of the propensity score in estimating dose-response
  functions}}.
\bjournal{Biometrika}
\bvolume{87}
\bpages{706--710}.
\end{barticle}
\endbibitem

\bibitem[\protect\citeauthoryear{Imbens}{2004}]{Imbens2004}
\begin{barticle}[author]
\bauthor{\bsnm{Imbens},~\bfnm{Guido~W.}\binits{G.~W.}}
(\byear{2004}).
\btitle{{Nonparametric estimation of average treatment effects under
  exogeneity: A review}}.
\bjournal{Review of Economics and Statistics}
\bvolume{86}
\bpages{4--29}.
\end{barticle}
\endbibitem

\bibitem[\protect\citeauthoryear{IOM}{2003}]{IOM2003}
\begin{bbook}[author]
\bauthor{\bsnm{IOM}}
(\byear{2003}).
\btitle{{Unequal Treatment: Confronting Racial and Ethnic Disparities in Health
  Care}}.
\bpublisher{The National Academies Press}, \baddress{Washington, DC}.
\end{bbook}
\endbibitem

\bibitem[\protect\citeauthoryear{J{\o}rgensen}{1997}]{Jorgensen2015}
\begin{bbook}[author]
\bauthor{\bsnm{J{\o}rgensen},~\bfnm{B.}\binits{B.}}
(\byear{1997}).
\btitle{{Theory of Dispersion Models}}.
\bpublisher{Chapman and Hall}, \baddress{London, UK}.
\end{bbook}
\endbibitem

\bibitem[\protect\citeauthoryear{Lechner}{2002}]{Lechner2002}
\begin{barticle}[author]
\bauthor{\bsnm{Lechner},~\bfnm{Michael}\binits{M.}}
(\byear{2002}).
\btitle{{Program heterogeneity and propensity score matching: An application to
  the evaluation of active labor market policies}}.
\bjournal{Review of Economics and Statistics}
\bvolume{84}
\bpages{205--220}.
\end{barticle}
\endbibitem

\bibitem[\protect\citeauthoryear{Li and Greene}{2013}]{LiGreene13}
\begin{barticle}[author]
\bauthor{\bsnm{Li},~\bfnm{Liang}\binits{L.}} \AND
  \bauthor{\bsnm{Greene},~\bfnm{Tom}\binits{T.}}
(\byear{2013}).
\btitle{A weighting analogue to pair matching in propensity score analysis}.
\bjournal{International Journal of Biostatistics}
\bvolume{9}
\bpages{1-20}.
\end{barticle}
\endbibitem

\bibitem[\protect\citeauthoryear{Li and Li}{2019a}]{LiLiSupp2019}
\begin{barticle}[author]
\bauthor{\bsnm{Li},~\bfnm{Fan}\binits{F.}} \AND
  \bauthor{\bsnm{Li},~\bfnm{Fan}\binits{F.}}
(\byear{2019}a).
\btitle{{Supplement to ``Propensity score weighting for causal inference with
  multiple treatments"}}.
\end{barticle}
\endbibitem

\bibitem[\protect\citeauthoryear{Li and Li}{2019b}]{LiLi2019}
\begin{barticle}[author]
\bauthor{\bsnm{Li},~\bfnm{Fan}\binits{F.}} \AND
  \bauthor{\bsnm{Li},~\bfnm{Fan}\binits{F.}}
(\byear{2019}b).
\btitle{{Double-robust estimation in difference-in-differences with an
  application to traffic safety evaluation}}.
\bjournal{Observational Studies}
\bvolume{5}
\bpages{1--20}.
\end{barticle}
\endbibitem

\bibitem[\protect\citeauthoryear{Li, Morgan and
  Zaslavsky}{2018}]{LiMorganZaslavsky2018}
\begin{barticle}[author]
\bauthor{\bsnm{Li},~\bfnm{Fan}\binits{F.}},
  \bauthor{\bsnm{Morgan},~\bfnm{Kari~Lock}\binits{K.~L.}} \AND
  \bauthor{\bsnm{Zaslavsky},~\bfnm{Alan~M.}\binits{A.~M.}}
(\byear{2018}).
\btitle{{Balancing covariates via propensity score weighting}}.
\bjournal{Journal of the American Statistical Association}
\bvolume{113}
\bpages{390--400}.
\end{barticle}
\endbibitem

\bibitem[\protect\citeauthoryear{Li, Thomas and Li}{2019}]{LiThomasLi2018}
\begin{barticle}[author]
\bauthor{\bsnm{Li},~\bfnm{Fan}\binits{F.}},
  \bauthor{\bsnm{Thomas},~\bfnm{Laine~E.}\binits{L.~E.}} \AND
  \bauthor{\bsnm{Li},~\bfnm{Fan}\binits{F.}}
(\byear{2019}).
\btitle{{Addressing extreme propensity scores via the overlap weights}}.
\bjournal{American Journal of Epidemiology}
\bvolume{1}
\bpages{250--257}.
\end{barticle}
\endbibitem

\bibitem[\protect\citeauthoryear{Li, Zaslavsky and
  Landrum}{2013}]{LiZaslavskyLandrum2013}
\begin{barticle}[author]
\bauthor{\bsnm{Li},~\bfnm{Fan}\binits{F.}},
  \bauthor{\bsnm{Zaslavsky},~\bfnm{Alan~M.}\binits{A.~M.}} \AND
  \bauthor{\bsnm{Landrum},~\bfnm{Mary~Beth}\binits{M.~B.}}
(\byear{2013}).
\btitle{{Propensity score weighting with multilevel data}}.
\bjournal{Statistics in Medicine}
\bvolume{32}
\bpages{3373--3387}.
\end{barticle}
\endbibitem

\bibitem[\protect\citeauthoryear{Lopez and Gutman}{2017}]{Lopez2017b}
\begin{barticle}[author]
\bauthor{\bsnm{Lopez},~\bfnm{Michael~J}\binits{M.~J.}} \AND
  \bauthor{\bsnm{Gutman},~\bfnm{Roee}\binits{R.}}
(\byear{2017}).
\btitle{{Estimation of causal effects with multiple treatments: A review and
  new ideas}}.
\bjournal{Statistical Science}
\bvolume{32}
\bpages{432--454}.
\end{barticle}
\endbibitem

\bibitem[\protect\citeauthoryear{Manning and Mullahy}{2001}]{Manning2001}
\begin{barticle}[author]
\bauthor{\bsnm{Manning},~\bfnm{W.~G.}\binits{W.~G.}} \AND
  \bauthor{\bsnm{Mullahy},~\bfnm{J.}\binits{J.}}
(\byear{2001}).
\btitle{{Estimating log models: to transform or not to transform?}}
\bjournal{Journal of Health Economics}
\bvolume{20}
\bpages{461--494}.
\end{barticle}
\endbibitem

\bibitem[\protect\citeauthoryear{McCaffrey et~al.}{2004}]{McCaffrey2004}
\begin{barticle}[author]
\bauthor{\bsnm{McCaffrey},~\bfnm{Daniel~F.}\binits{D.~F.}},
  \bauthor{\bsnm{Ridgeway},~\bfnm{G.}\binits{G.}}, \bauthor{} \AND
  \bauthor{\bsnm{Morral},~\bfnm{A.}\binits{A.}}
(\byear{2004}).
\btitle{{Propensity score estimation with boosted regression for evaluating
  causal effects in observational studies}}.
\bjournal{Psychological Methods}
\bvolume{9}
\bpages{403--425}.
\end{barticle}
\endbibitem

\bibitem[\protect\citeauthoryear{McCaffrey et~al.}{2013}]{McCaffrey2013}
\begin{barticle}[author]
\bauthor{\bsnm{McCaffrey},~\bfnm{Daniel~F.}\binits{D.~F.}},
  \bauthor{\bsnm{Griffin},~\bfnm{Beth~Ann}\binits{B.~A.}},
  \bauthor{\bsnm{Almirall},~\bfnm{Daniel}\binits{D.}},
  \bauthor{\bsnm{Slaughter},~\bfnm{Mary~Ellen}\binits{M.~E.}},
  \bauthor{\bsnm{Ramchand},~\bfnm{Rajeev}\binits{R.}} \AND
  \bauthor{\bsnm{Burgette},~\bfnm{Lane~F.}\binits{L.~F.}}
(\byear{2013}).
\btitle{{A tutorial on propensity score estimation for multiple treatments
  using generalized boosted models}}.
\bjournal{Statistics in Medicine}
\bvolume{32}
\bpages{3388--3414}.
\end{barticle}
\endbibitem

\bibitem[\protect\citeauthoryear{McGuire et~al.}{2006}]{McGuire2006}
\begin{barticle}[author]
\bauthor{\bsnm{McGuire},~\bfnm{Thomas~G.}\binits{T.~G.}},
  \bauthor{\bsnm{Alegria},~\bfnm{Margarita}\binits{M.}},
  \bauthor{\bsnm{Cook},~\bfnm{Benjamin~L.}\binits{B.~L.}},
  \bauthor{\bsnm{Wells},~\bfnm{Kenneth~B.}\binits{K.~B.}} \AND
  \bauthor{\bsnm{Zaslavsky},~\bfnm{Alan~M.}\binits{A.~M.}}
(\byear{2006}).
\btitle{{Implementing the Institute of Medicine definition of disparities: An
  application to mental health care}}.
\bjournal{Health Services Research}
\bvolume{41}
\bpages{1979--2005}.
\end{barticle}
\endbibitem

\bibitem[\protect\citeauthoryear{Moore et~al.}{2012}]{Moore2012}
\begin{barticle}[author]
\bauthor{\bsnm{Moore},~\bfnm{Kelly~L.}\binits{K.~L.}},
  \bauthor{\bsnm{Neugebauer},~\bfnm{Romain}\binits{R.}}, \bauthor{\bsnm{{Van
  der Laan}},~\bfnm{Mark~J.}\binits{M.~J.}} \AND
  \bauthor{\bsnm{Tager},~\bfnm{Ira~B.}\binits{I.~B.}}
(\byear{2012}).
\btitle{{Causal inference in epidemiological studies with strong confounding}}.
\bjournal{Statistics in Medicine}
\bvolume{31}
\bpages{1380--1404}.
\end{barticle}
\endbibitem

\bibitem[\protect\citeauthoryear{Mu{\~{n}}oz and van~der
  Laan}{2012}]{Munoz2012}
\begin{barticle}[author]
\bauthor{\bsnm{Mu{\~{n}}oz},~\bfnm{Iv{\'{a}}n~D{\'{i}}az}\binits{I.~D.}} \AND
  \bauthor{\bparticle{van~der} \bsnm{Laan},~\bfnm{Mark}\binits{M.}}
(\byear{2012}).
\btitle{{Population Intervention Causal Effects Based on Stochastic
  Interventions}}.
\bjournal{Biometrics}
\bvolume{68}
\bpages{541--549}.
\end{barticle}
\endbibitem

\bibitem[\protect\citeauthoryear{Park}{1966}]{Park1966}
\begin{barticle}[author]
\bauthor{\bsnm{Park},~\bfnm{R.}\binits{R.}}
(\byear{1966}).
\btitle{Estimation with heteroscedastic error terms}.
\bjournal{Econometrica}
\bvolume{34}
\bpages{888}.
\end{barticle}
\endbibitem

\bibitem[\protect\citeauthoryear{Pirracchio, Petersen and van~der
  Laan}{2015}]{Pirracchio2015}
\begin{barticle}[author]
\bauthor{\bsnm{Pirracchio},~\bfnm{Romain}\binits{R.}},
  \bauthor{\bsnm{Petersen},~\bfnm{Maya~L.}\binits{M.~L.}} \AND
  \bauthor{\bparticle{van~der} \bsnm{Laan},~\bfnm{Mark}\binits{M.}}
(\byear{2015}).
\btitle{{Improving propensity score estimators' robustness to model
  misspecification using Super Learner}}.
\bjournal{American Journal of Epidemiology}
\bvolume{181}
\bpages{108--119}.
\end{barticle}
\endbibitem

\bibitem[\protect\citeauthoryear{Rassen et~al.}{2013}]{Rassen2013}
\begin{barticle}[author]
\bauthor{\bsnm{Rassen},~\bfnm{Jeremy~A.}\binits{J.~A.}},
  \bauthor{\bsnm{Shelat},~\bfnm{Abhi~A.}\binits{A.~A.}},
  \bauthor{\bsnm{Franklin},~\bfnm{Jessica~M.}\binits{J.~M.}},
  \bauthor{\bsnm{Glynn},~\bfnm{Robert~J.}\binits{R.~J.}},
  \bauthor{\bsnm{Solomon},~\bfnm{Daniel~H.}\binits{D.~H.}} \AND
  \bauthor{\bsnm{Schneeweiss},~\bfnm{Sebastian}\binits{S.}}
(\byear{2013}).
\btitle{{Matching by propensity score in cohort studies with three treatment
  groups}}.
\bjournal{Epidemiology}
\bvolume{24}
\bpages{401--409}.
\end{barticle}
\endbibitem

\bibitem[\protect\citeauthoryear{Robins, Rotnitzky and Zhao}{1994}]{Robins1994}
\begin{barticle}[author]
\bauthor{\bsnm{Robins},~\bfnm{J~M}\binits{J.~M.}},
  \bauthor{\bsnm{Rotnitzky},~\bfnm{A}\binits{A.}} \AND
  \bauthor{\bsnm{Zhao},~\bfnm{L~P}\binits{L.~P.}}
(\byear{1994}).
\btitle{{Estimation of regression-coefficients when some regressors are not
  always observed}}.
\bjournal{Journal of the American Statistical Association}
\bvolume{89}
\bpages{846--866}.
\end{barticle}
\endbibitem

\bibitem[\protect\citeauthoryear{Robins et~al.}{2008}]{Robins2008}
\begin{barticle}[author]
\bauthor{\bsnm{Robins},~\bfnm{James}\binits{J.}},
  \bauthor{\bsnm{Li},~\bfnm{Lingling}\binits{L.}},
  \bauthor{\bsnm{Tchetgen},~\bfnm{Eric~Tchetgen}\binits{E.~T.}} \AND
  \bauthor{\bparticle{van~der} \bsnm{Vaart},~\bfnm{Aad}\binits{A.}}
(\byear{2008}).
\btitle{{Higher order influence functions and minimax estimation of nonlinear
  functionals}}.
\bjournal{Institute of Mathematical Statistics Collections. Probability and
  Statistics: Essays in Honor of David A. Freedman}
\bvolume{2}
\bpages{335--421}.
\end{barticle}
\endbibitem

\bibitem[\protect\citeauthoryear{Rosenbaum and Rubin}{1983}]{Rosenbaum1983}
\begin{barticle}[author]
\bauthor{\bsnm{Rosenbaum},~\bfnm{Paul~R.}\binits{P.~R.}} \AND
  \bauthor{\bsnm{Rubin},~\bfnm{Donald~B.}\binits{D.~B.}}
(\byear{1983}).
\btitle{{The central role of the propensity score in observational studies for
  causal effects}}.
\bjournal{Biometrika}
\bvolume{70}
\bpages{41--55}.
\end{barticle}
\endbibitem

\bibitem[\protect\citeauthoryear{Stefanski and Boos}{2002}]{Stefanski2002}
\begin{barticle}[author]
\bauthor{\bsnm{Stefanski},~\bfnm{L.~A.}\binits{L.~A.}} \AND
  \bauthor{\bsnm{Boos},~\bfnm{D.~D}\binits{D.~D.}}
(\byear{2002}).
\btitle{{The calculus of M-estimation}}.
\bjournal{American Statistician}
\bvolume{56}
\bpages{29--38}.
\end{barticle}
\endbibitem

\bibitem[\protect\citeauthoryear{van~der Laan and
  Petersen}{2007}]{VanderLaan2007}
\begin{barticle}[author]
\bauthor{\bparticle{van~der} \bsnm{Laan},~\bfnm{Mark~J.}\binits{M.~J.}} \AND
  \bauthor{\bsnm{Petersen},~\bfnm{Maya~L.}\binits{M.~L.}}
(\byear{2007}).
\btitle{{Causal effect models for realistic individualized treatment and
  intention to treat rules}}.
\bjournal{International Journal of Biostatistics}
\bvolume{3}
\bpages{1--51}.
\end{barticle}
\endbibitem

\bibitem[\protect\citeauthoryear{VanderWeele and
  Robinson}{2014a}]{VanderWeele2014a}
\begin{barticle}[author]
\bauthor{\bsnm{VanderWeele},~\bfnm{Tyler~J.}\binits{T.~J.}} \AND
  \bauthor{\bsnm{Robinson},~\bfnm{Whitney~R.}\binits{W.~R.}}
(\byear{2014}a).
\btitle{{On the causal interpretation of race in regressions adjusting for
  confounding and mediating variables}}.
\bjournal{Epidemiology}
\bvolume{25}
\bpages{473--484}.
\end{barticle}
\endbibitem

\bibitem[\protect\citeauthoryear{VanderWeele and
  Robinson}{2014b}]{VanderWeele2014b}
\begin{barticle}[author]
\bauthor{\bsnm{VanderWeele},~\bfnm{Tyler~J.}\binits{T.~J.}} \AND
  \bauthor{\bsnm{Robinson},~\bfnm{Whitney~R.}\binits{W.~R.}}
(\byear{2014}b).
\btitle{{Rejoinder: How to reduce racial disparities?: Upon what to
  intervene?}}
\bjournal{Epidemiology}
\bvolume{25}
\bpages{491--493}.
\end{barticle}
\endbibitem

\bibitem[\protect\citeauthoryear{Yang et~al.}{2016}]{Yang2016}
\begin{barticle}[author]
\bauthor{\bsnm{Yang},~\bfnm{Shu}\binits{S.}},
  \bauthor{\bsnm{Imbens},~\bfnm{Guido~W.}\binits{G.~W.}},
  \bauthor{\bsnm{Cui},~\bfnm{Zhanglin}\binits{Z.}},
  \bauthor{\bsnm{Faries},~\bfnm{Douglas~E.}\binits{D.~E.}} \AND
  \bauthor{\bsnm{Kadziola},~\bfnm{Zbigniew}\binits{Z.}}
(\byear{2016}).
\btitle{{Propensity score matching and subclassification in observational
  studies with multi-level treatments}}.
\bjournal{Biometrics}
\bvolume{72}
\bpages{1055--1065}.
\end{barticle}
\endbibitem

\bibitem[\protect\citeauthoryear{Yoshida et~al.}{2017}]{Yoshida2017}
\begin{barticle}[author]
\bauthor{\bsnm{Yoshida},~\bfnm{Kazuki}\binits{K.}},
  \bauthor{\bsnm{Hern{\'{a}}ndez-D{\'{i}}az},~\bfnm{Sonia}\binits{S.}},
  \bauthor{\bsnm{Solomon},~\bfnm{Daniel~H.}\binits{D.~H.}},
  \bauthor{\bsnm{Jackson},~\bfnm{John~W.}\binits{J.~W.}},
  \bauthor{\bsnm{Gagne},~\bfnm{Joshua~J.}\binits{J.~J.}},
  \bauthor{\bsnm{Glynn},~\bfnm{Robert~J.}\binits{R.~J.}} \AND
  \bauthor{\bsnm{Franklin},~\bfnm{Jessica~M.}\binits{J.~M.}}
(\byear{2017}).
\btitle{{Matching weights to simultaneously compare three treatment groups
  comparison to three-way matching}}.
\bjournal{Epidemiology}
\bvolume{28}
\bpages{387--395}.
\end{barticle}
\endbibitem

\bibitem[\protect\citeauthoryear{Zanutto, Lu and Hornik}{2005}]{Zanutto2005}
\begin{barticle}[author]
\bauthor{\bsnm{Zanutto},~\bfnm{E.}\binits{E.}},
  \bauthor{\bsnm{Lu},~\bfnm{B.}\binits{B.}} \AND
  \bauthor{\bsnm{Hornik},~\bfnm{R.}\binits{R.}}
(\byear{2005}).
\btitle{{Using propensity score subclassification for multiple treatment doses
  to evaluate a national antidrug media campaign}}.
\bjournal{Journal of Educational and Behavioral Statistics}
\bvolume{30}
\bpages{59--73}.
\end{barticle}
\endbibitem

\bibitem[\protect\citeauthoryear{Zaslavsky and Ayanian}{2005}]{Zaslavsky2005}
\begin{barticle}[author]
\bauthor{\bsnm{Zaslavsky},~\bfnm{Alan~M}\binits{A.~M.}} \AND
  \bauthor{\bsnm{Ayanian},~\bfnm{John~Z}\binits{J.~Z.}}
(\byear{2005}).
\btitle{{Integrating research on racial and ethnic disparities in health care
  over place and time}}.
\bjournal{Medical Care}
\bvolume{43}
\bpages{303--307}.
\end{barticle}
\endbibitem

\bibitem[\protect\citeauthoryear{Zubizarreta}{2015}]{Zubizarreta2015}
\begin{barticle}[author]
\bauthor{\bsnm{Zubizarreta},~\bfnm{Jos{\'{e}}~R.}\binits{J.~R.}}
(\byear{2015}).
\btitle{{Stable Weights that Balance Covariates for Estimation With Incomplete
  Outcome Data}}.
\bjournal{Journal of the American Statistical Association}
\bvolume{110}
\bpages{910--922}.
\bdoi{10.1080/01621459.2015.1023805}
\end{barticle}
\endbibitem

\end{thebibliography}

\end{document}